\documentclass[prb,twocolumn,showpacs,nofootinbib,preprintnumbers,amsmath,amssymb,aps,longbibliography,superscriptaddress]{revtex4-2}
\usepackage[T1]{fontenc}
\usepackage[utf8]{inputenc}
\usepackage{subfigure, lmodern, amsmath,amssymb, graphicx, pifont, adjustbox, bm, xcolor}
\usepackage{amsfonts}
\usepackage{amsthm}
\usepackage{comment}
\usepackage{mathtools}
\usepackage{float}
\usepackage{braket}
\usepackage{longtable}
\usepackage{graphicx, subfigure}
\usepackage{xcolor}
\usepackage{tikz}
\usetikzlibrary{calc,decorations.markings,arrows.meta}
\usepackage{enumitem}
\usepackage{dsfont}
\usepackage{multirow}
\usepackage{pifont}
\usepackage[normalem]{ulem}

\usepackage[colorlinks=true, citecolor=blue, urlcolor=blue, linkcolor=blue, breaklinks=true, pdfpagelabels=false]{hyperref}

\newcommand{\appendixref}[1]{\hyperref[#1]{appendix~\ref{#1}}}
\def\equationautorefname~#1\null{eq.\,(#1)\null}

\usepackage{cleveref}
\crefname{appendix}{App.}{Apps.}
\crefname{equation}{Eq.}{Eqs.}
\crefname{figure}{Fig.}{Figs.}
\crefname{table}{Tab.}{Tabs.}
\crefname{section}{Sec.}{Secs.}
\crefname{theorem}{Theorem}{Theorem}
\crefname{definition}{Definition}{Definition}
\crefname{lemma}{Lemma}{Lemma}

\newcommand{\hQ}{\hat{Q}}
\newcommand{\hbroken}{\hat{\Gamma}}
\newcommand{\hH}{{H}}
\newcommand{\hP}{\hat{P}}
\newcommand{\hU}{\hat{U}}
\newcommand{\hT}{\hat{T}}
\newcommand{\hn}{\hat{n}}
\newcommand{\hb}{\hat{b}}

\newcommand{\goldstone}{\varphi}

\newcommand{\GS}{\Omega}
\newcommand{\lket}[1]{\langle #1 |}
\newcommand{\rket}[1]{|#1\rangle}
\newcommand{\lrket}[2]{\langle #1 |#2\rangle}
\newcommand{\lrangle}[1]{\langle #1 \rangle}

\newcommand{\hQexp}{\hQ_\text{e}}
\newcommand{\hQsin}{\hQ_\text{s}}
\newcommand{\hQcos}{\hQ_\text{c}}
\newcommand{\hQset}{\hat{\bold{Q}}}
\newcommand{\hQmul}[1]{\hQ_{#1}}
\newcommand{\hj}{\hat{j}}
\newcommand{\diffD}{\text{d}}

\newcommand{\qSet}{\textbf{q}}

\newcommand{\orderP}[1]{\hat{\Phi}_{#1}}

\newcommand{\goldstoneSet}{\boldsymbol{\varphi}}
\newcommand{\boldx}{\boldsymbol{x}}
\newcommand{\JordanReal}{\mathcal{J}^{\text{R}}}
\newcommand{\JordanComplex}{\mathcal{J}^{\text{C}}}
\newcommand{\boldj}{\hat{\boldsymbol{J}}}
\newcommand{\modulatingFunction}[1]{F(#1)}
\newcommand{\modulatingFunctionIndex}[2]{f_{#1}(#2)}

\newcommand{\cmark}{\ding{51}}
\newcommand{\xmark}{\ding{55}}

\begin{document}

\title{Translationally Covariant Modulated Symmetries: Classification and Goldstone}
\date{\today}

\author{Bo-Ting Chen}
\affiliation{Department of Physics, Princeton University, Princeton, New Jersey 08544, USA}
\author{Zihan Zhou}
\affiliation{Department of Physics, Princeton University, Princeton, New Jersey 08544, USA}
\author{Biao Lian}
\affiliation{Department of Physics, Princeton University, Princeton, New Jersey 08544, USA}

\begin{abstract}
Modulated symmetries are global symmetries with a spatially dependent unit of charge, such as the dipole symmetry and the exponential symmetry. We give the generic condition for a modulated symmetry to be compatible with translationally symmetric Hamiltonians, which we define as a translationally covariant modulated symmetry (TCMS). For Abelian TCMSs, we prove that their units of charge can only contain multipole, exponential and harmonic components. Particularly, we classify all the one-dimensional TCMSs by real Jordan normal form blocks. We further derive the generic Goldstone action for SSB phases of continuous TCMSs, by which we show that a broken multipole symmetry gives higher-order gapless Goldstone modes, a broken harmonic symmetry gives gapless Goldstone modes at finite momenta, and a broken exponential symmetry gives no gapless Goldstone modes, modifying the conventional Goldstone theorem.
\end{abstract}

\maketitle

\emph{Introduction.} Symmetries play a central role in the characterization of both conventional and topological phases of matter. Specifically, spontaneous symmetry breaking (SSB) serves as a significant mechanism for a large class of ordered phases and phase transitions, and the Goldstone theorem \cite{Nambu1960,Goldstone1961,Watanabe_2012,Hidaka_2013,Watanabe_2013,Hidaka_2015,Watanabe_2020}
ensures the existence of gapless Goldstone modes in the SSB of translationally invariant continuous symmetries. Recently, generalized symmetries of various types have attracted extensive interests of study \cite{Gaiotto_2015,Ji_2020,Chatterjee:2022jll,McGreevy_2023,cordova2022snowmasswhitepapergeneralized,bhardwaj2023lecturesgeneralizedsymmetries,schafernameki2023ictplecturesnoninvertiblegeneralized},
for their promise to protect novel unconventional phases in quantum many-body systems. A particular class of on-site symmetries are denoted as \emph{modulated symmetries} \cite{Sala_2022_dynamics}, which are defined to have the unit of symmetry charge (if Abelian) be a spatially non-uniform function $\modulatingFunction{\boldx}$ of site position $\boldx$. The modulated symmetries studied so far include the dipole symmetry, multipole symmetry, exponential symmetry, and various subsystem symmetries, which can lead to strongly constrained dynamics, fractonic phases \cite{Gorantla_2022,Delfino_2023rpw,Gorantla_2021,Seiberg_2021,grosvenor2021spacedependentsymmetriesfractons,Hirono_2024}, unconventional SSB phases and symmetry protected topological (SPT) phases \cite{Han_2023fas,Lam_classification,Saito_2025,Kim_2025,Yao:2025iia,ning2026matrixproductstatesmodulated,anakru2026matrixproductstatesmodulated}.

Evidently, with a spatially varying $\modulatingFunction{\boldx}$, the modulated symmetry charge does not commute with spatial translation. However, the Hamiltonian $H$ with a modulated symmetry can still have translation symmetry, such as the dipole conserving model \cite{Bergholtz_2008,Nakamura:2012ujh,Schulz_2019,Yuan_2020,Chen_2021,Sala_2020,Lake_2022,Khemani_2020,Pai_2019}
and the quantum breakdown model \cite{Lian_2023,Chen_2024_EQBM,Liu_2025,hu2024bosonicquantumbreakdownhubbard,Hu_2024_lattice,hu2025quantumbreakdowncondensatedisorderfree}. Moreover, these models can have SSB phases with uniform particle density and unconventional Goldstone properties (for continuous modulated symmetries). For instance, the SSB of dipole and multipole conserving models have higher order Goldstone mode dispersions \cite{Gromov_2019,Stahl_2022,Lake_2023}, while the SSB of quantum breakdown model with exponential symmetry is shown to have no gapless Goldstone modes \cite{hu2024bosonicquantumbreakdownhubbard,hu2025quantumbreakdowncondensatedisorderfree}.

This motivates us to study two questions for generic continuous modulated symmetries. First, what modulated symmetries are compatible with translationally symmetric Hamiltonians? We define such symmetries as \emph{translationally covariant modulated symmetries}. Second, do their SSB have universal properties of Goldstone modes? In this letter, we derive the most generic form of continuous Abelian translationally covariant symmetries, by which we prove that their unit-of-charge functions $\modulatingFunction{\boldx}$ can only contain polynomial (i.e. multipole), exponential and harmonic components. We further derive the universal effective Goldstone action in their SSB, which is independent of microscopic models. Particularly, the gapless Goldstone modes are absent if and only if the broken symmetries contain exponential components.

\emph{General formulation}. We restrict ourselves to continuous modulated symmetries in a $d$-dimensional infinite flat space. We assume that a modulated symmetry has a complete set of generators given by $N$ linearly independent modulated symmetry charges $\hat Q^{(m)}$ ($1\le m\le N$), which are summations of local operators. We write these charges in vector form as $\hQset = (\hQ^{(1)},\dots,\hQ^{(N)})^{T}$. 

If there exists a translationally invariant Hamiltonian $H$ that commutes with the above charges $\hQset$ and has no other on-site symmetry, we define the symmetry generated by $\hQset$ as a \emph{translationally covariant modulated symmetry} (TCMS). For generality, we consider a continuous flat space with $\hat P_\mu$ being the translation generator (momentum operator) in the $\mu$-th dimension in Cartesian coordinates ($1\le \mu\le d$). The reduction to lattices can be done by restricting the translation to lattice vectors. The sufficient and necessary condition for a TCMS is that, the symmetry charges $\hQset$ form a closed set under translation. This requires the following adjoint action of translation:
\begin{equation}
\label{eqn: adjoint action of P}
    \mathrm{ad}_{\hP_\mu}(\hQset)
    \equiv
    [\hP_\mu,\, \hQset]
    = -i A_{\mu}\, \hQset\ ,
\end{equation}
where $A_\mu$ are $N\times N$ structure constant matrices. The Hermiticity requires $A_\mu$ to be real. The translation by an infinitesimal $\delta x^\mu$ transforms $\hQset$ to $\hQset-i[\hP_\mu,\, \hQset]\delta x^\mu$, which preserves locality, thus $[\hP_\mu,\, \hQset]$ must be linear in $\hQset$ and independent of $\hP_\mu$. By the Jacobi identity and noting that $[\hP_\mu,\hP_\nu]=0$, we have $0=[\hQset,[\hP_\mu,\hP_\nu]]=[\hP_\nu,[\hP_\mu,\hQset]]-[\hP_\mu,[\hP_\nu,\hQset]]=[A_\mu,A_\nu]\hQset$, so $A_\mu$ must satisfy
\begin{equation}
\label{eqn: commuting A}
[A_\mu,A_\nu]=0\ .
\end{equation}
\cref{eqn: adjoint action of P,eqn: commuting A} define the most generic TCMS. Besides, the commutators between different charges are of the form $[\hQ^{(m)},\hQ^{(n)}]=i B^{mn}_\ell \hQ^{(\ell)}$, with the structure constants $B^{mn}_\ell$ subject to Jacobi identity constraints (see End Matter).

Hereafter, we will only consider Abelian TCMSs, which are defined to have all $B^{mn}_\ell=0$. 
The modulated symmetries of all known models so far belong to this class. Note that by \cref{eqn: adjoint action of P}, $\hP_\mu$ and $\hQset$ together can generate a non-Abelian symmetry group, so `Abelian' here solely refers to a property within $\hQset$.

In terms of spatial coordinate $\boldx=(x^1,\cdots,x^d)$, the generic Abelian TCMS charges are given by the solution to \cref{eqn: adjoint action of P}, which is of the form
\begin{equation}\label{eqn:Qsolution}
\hQset=\int e^{A_\mu x^\mu}\boldj_0(\boldx) d^d\boldx\ ,
\end{equation}
where (and hereafter) we assumed Einstein summation over repeated indices. $\boldj_0=\big(\hat J_0^{(1)},\cdots,\hat J_0^{(N)}\big)^T$ are $N$ commuting local operators satisfying $[\hP_\mu,\boldj_0]=i\partial_\mu \boldj_0$, which ensures that $e^{-i\hP_\mu y^\mu}\boldj_0(\boldx)e^{i\hP_\mu y^\mu}=\boldj_0(\boldx+\boldsymbol{y})$ under translation by $\boldsymbol{y}$. \cref{eqn:Qsolution} also applies to lattice models by reducing the integral in $\boldx$ into a sum in $\boldx$ running over all the lattice sites. Each charge $\hQ^{(m)}$ generates either a compact U$(1)$ or non-compact $\mathbb{R}$ symmetry. \cref{eqn:Qsolution} shows that the TCMS has a position dependent matrix-form unit of charge $\modulatingFunction{\boldx}=e^{A_\mu x^\mu}$. Note that this resembles a Wilson line from $\boldsymbol{0}$ to $\boldx$, in which $-iA_\mu$ plays the role of a constant imaginary connection.

By \cref{eqn: commuting A}, all the matrices $A_\mu$ ($\mu=1,\cdots,d$) commute, so they can be simultaneously transformed to \emph{upper triangular} by a similarity transformation $A_\mu\rightarrow S A_\mu S^{-1}$ with a complex matrix $S$, which corresponds to a linear recombination of charges $\hQset\rightarrow S\hQset$. In the upper triangular form of $A_\mu$, it is instructive to understand two special cases:

(i) If $A_\mu$ has all diagonal elements zero, it is nilpotent, namely $(A_\mu)^p=0$ for some integer $p\ge 1$. Therefore, by Taylor expansion, the unit of charge $\modulatingFunction{\boldx}=e^{A_\mu x^\mu}$ contains only polynomials of $x^\mu$ with degree no higher than $(p-1)$. This gives a multipole symmetry up to the $(p-1)$-th multipole moment in the $x^\mu$ direction.

(ii) If $A_\mu$ is diagonal, the unit of charge $\modulatingFunction{\boldx}=e^{A_\mu x^\mu}$ will exponentially grow or decay (harmonically oscillate) in $x_\mu$ if the eigenvalues of $A_\mu$ have nonzero real (imaginary) parts. This gives an exponential or harmonic symmetry in the $x^\mu$ direction.

Generically, a TCMS charge may simultaneously contain multipole, exponential, and harmonic components, which may depend on spatial directions $x^\mu$. Specifically, if all matrices $A_\mu$ are proportional to each other, i.e., $A_\mu=\zeta_\mu A$, one can rotate the coordinates to align vector $\boldsymbol{\zeta}=(\zeta_1,\cdots,\zeta_d)^T$ with the $x^1$ axis. After the rotation, such a TCMS will have all $A_{\mu\ge 2}=0$ and only $A_1=|\boldsymbol{\zeta}|A$ nonzero, which we denote as a \emph{1D TCMS}.

\emph{1D TCMS classification}. 
We can more explicitly classify the 1D TCMSs, which have only $A_1$ nonzero. By a similarity transformation $A_1\rightarrow S A_1 S^{-1}$ with a real matrix $S$, one can bring $A_1$ to a block-diagonal real Jordan normal form as follows:
\begin{equation}
\begin{split}
A_1=\text{diag}\Big(
& \JordanReal_{p_1}(\alpha_1),\cdots,\JordanReal_{p_r}(\alpha_r), \\
        & 
        \JordanComplex_{l_1}(\alpha_1',\beta_1'),\cdots,
        \JordanComplex_{l_s}(\alpha_s',\beta_s')
\Big)\ ,
\end{split}
\end{equation}
where $p_j$ and $l_j$ are positive integers satisfying $\sum_{j=1}^r p_j +2\sum_{j=1}^{s}l_j=N$. Here $\JordanReal_{p}(\alpha)$ is a $p\times p$ Jordan block associated with a real eigenvalue $\alpha\in\mathbb{R}$:
\begin{equation}\label{eqn:Jreal}
    \JordanReal_p(\alpha) = \begin{pmatrix}
        \alpha & 1 & & 0 \\
        & \alpha & \ddots & \\
        & & \ddots & 1 \\
        0 & & & \alpha
    \end{pmatrix},
\end{equation}
and $\JordanComplex_{l}(\alpha,\beta)$ is a $2l\times 2l$ Jordan block associated with a complex conjugate pair of eigenvalues $\alpha\pm i\beta$ ($\alpha,\beta\in\mathbb{R}$):
\begin{equation}\label{eqn:Jcomplex}
    \JordanComplex_l(\alpha,\beta)
    = \begin{pmatrix}
        C_{\alpha,\beta} & I_2 & & 0 \\
        & C_{\alpha,\beta} & \ddots & \\
        & & \ddots & I_2 \\
        0 & & & C_{\alpha,\beta}
    \end{pmatrix},\ C_{\alpha,\beta}
=\begin{pmatrix}
    \alpha & \beta \\
    -\beta & \alpha
\end{pmatrix}.
\end{equation}
Each Jordan block forms a subgroup of the TCMS, which we denote as an \emph{irreducible} TCMS. This classifies all the 1D TCMSs. Equivalently, this classification can be derived from a differential equation method given in the SM \cite{SM}.
We specifically name the following three simple families of irreducible TCMSs:

(i) $A_1=\JordanReal_p(0)$ (with $N=p$), which gives a \emph{multipole symmetry}. By \cref{eqn:Jreal}, $A_1$ is nilpotent with $(A_1)^p=0$. Thus, the matrix $\modulatingFunction{\boldx}=e^{A_1 x^1}$ only consists of polynomials of $x^1$ up to degree $(p-1)$, which yields a multipole symmetry up to the $(p-1)$-th moment. As a simple example, if $\boldj_0(\boldx)=\big(0,\cdots,0,\hat n(\boldx)\big)^T$ where $\hat n(\boldx)$ is the particle density, one has $\hQ^{(m)}=\int \frac{(x^1)^{p-m}}{(p-m)!}\hat n(\boldx) d^d\boldx$ being the $(p-m)$-th multipole moment ($1\le m\le p$). Note that a multipole symmetry necessarily has $p$ linearly independent charges of moments from $0$ to $p-1$.

(ii) $A_1=\JordanReal_p(\alpha)$ with $\alpha\neq 0$ ($N=p$), which gives an \emph{exponential symmetry}. If $p=1$, one arrives at a pristine exponential symmetry with an exponential unit of charge $\modulatingFunction{\boldx}=e^{\alpha x^1}$. If $p\ge 2$, this gives a dressed exponential symmetry with $\modulatingFunction{\boldx}$ matrix elements given by $e^{\alpha x^1}$ times polynomials of degree no higher than $p-1$.

(iii) $A_1=\JordanComplex_l(0,\beta)$ with $\beta>0$ ($N=2l$), which we denote as a \emph{harmonic symmetry}. If $l=1$, one obtains a pristine harmonic symmetry with $\modulatingFunction{\boldx}$ being a $2\times2$ matrix consisting of harmonic functions $\sin(\beta x^1)$ and $\cos(\beta x^1)$. For example, $\boldj_0(\boldx)=\big(0,\hat n(\boldx)\big)^T$ gives two harmonic symmetry charges $\hQ^{(1)}=\hQsin=\int \sin(\beta x^1)\hat n(\boldx)d^d\boldx$ and $\hQ^{(2)}=\hQcos=\int \cos(\beta x^1)\hat n(\boldx)d^d\boldx$. If $l\ge 2$, one obtains a dressed harmonic symmetry with $\modulatingFunction{\boldx}$ being harmonic functions times polynomials of degree no higher than $l-1$.

The following translationally invariant 1D deformed Bose-Hubbard models are examples with a 1D TCMS:
\begin{equation}\label{eqn:boson-model}
H= \sum_{j} \left[ V(\hn_{j}) - J \left(\hat{K}_{j} + \text{h.c.}\right) \right]\ ,
\end{equation}
where $j\in\mathbb{Z}$ labels the 1D lattice site, $\hat n_j=\hat b_j^\dag \hat b_j$ is the particle number on site $j$, the on-site potential $V(\hn_{j})$ is a lower-bounded function of $\hn_{j}$, and $\hat{K}_{j}$ is a product of the boson creation or annihilation operators around site $j$. \cref{table: various models} shows the operator $\hat{K}_{j}$ for three models in the literature with dipole symmetry \cite{Lake_2022}, exponential symmetry \cite{hu2024bosonicquantumbreakdownhubbard} and harmonic symmetry \cite{Sala_2024_exotic}, respectively, in which $V(\hn_{j})=\frac{U}{2} \hn_{j} (\hn_{j}-1)+\frac{U'}{6} \hn_{j} (\hn_{j}-1)(\hn_{j}-2)-\mu \hn_{j}$. We set $U>0$ in general, and $U'>0$ only in the dipole symmetry Hamiltonian to ensure it is lower bounded. 
The modulated symmetry charges in \cref{eqn:Qsolution} are reduced to lattice sites $x^1=j\in\mathbb{Z}$,  as shown in \cref{table: various models}, where the structure constant matrix is $A_1=\JordanReal_2(0)$ for the dipole model, $A_1=\JordanReal_1(-\ln 2)$ for the exponential model (known as the quantum breakdown model \cite{hu2024bosonicquantumbreakdownhubbard}), and $A_1=\JordanComplex_1(0,\frac{\pi}{3})$ for the harmonic model, respectively. These models with 1D TCMS can be easily generalized to $d\ge 2$ dimensions by adding normal hoppings in the $x^\mu$ directions with $2\le\mu\le d$. They can also be mapped to spin models by replacing $\hat b_j$ and $\hat b_j^\dag$ with the spin raising and lowering operators $\hat S_j^\pm$ \cite{Sala_2020,Khemani_2020,Khudorozhkov_2022,Srivastava_2025,hu2025quantumbreakdowncondensatedisorderfree}.

\begin{table}[t]
\centering
\caption{The interaction of three 1D deformed Bose-Hubbard models and their corresponding TCMSs and conserved charges.}
\label{table: various models}
\begin{tabular}{|c|c|l|}
\hline
$\hat{K}_{j}$ & TCMS & modulated symmetry charges \\
\hline
\hline
\multirow{2}{*}{$\hb_{j-1}^{\dagger} \hb_{j}^{2} \hb_{j+1}^{\dagger}$} & \multirow{2}{*}{dipole} & $\hQmul{\text{d}} = \sum_j j \hn_j$\quad (dipole) \\
\cline{3-3}
 &  & $\hQmul{\text{m}} = \sum_j \hn_j$\quad (monopole) \\
\hline
$\hb_{j} \hb_{j+1}^{\dagger 2}$ & exponential & $\hQexp = \sum_j 2^{-j} \hn_j$\quad (exponential) \\
\hline
\multirow{2}{*}{$\hb_{j-1}^{\dagger} \hb_{j} \hb_{j+1}^{\dagger}$} & \multirow{2}{*}{harmonic} & $\hQsin = \sum_j \sin\left( \frac{j\pi}{3} \right) \hn_j$\quad (sine) \\
\cline{3-3}
 & & $\hQcos = \sum_j \cos\left( \frac{j\pi}{3} \right) \hn_j$\quad (cosine) \\
\hline
\end{tabular}
\end{table}

\begin{figure}[t]
\centering
\includegraphics[width=90mm]{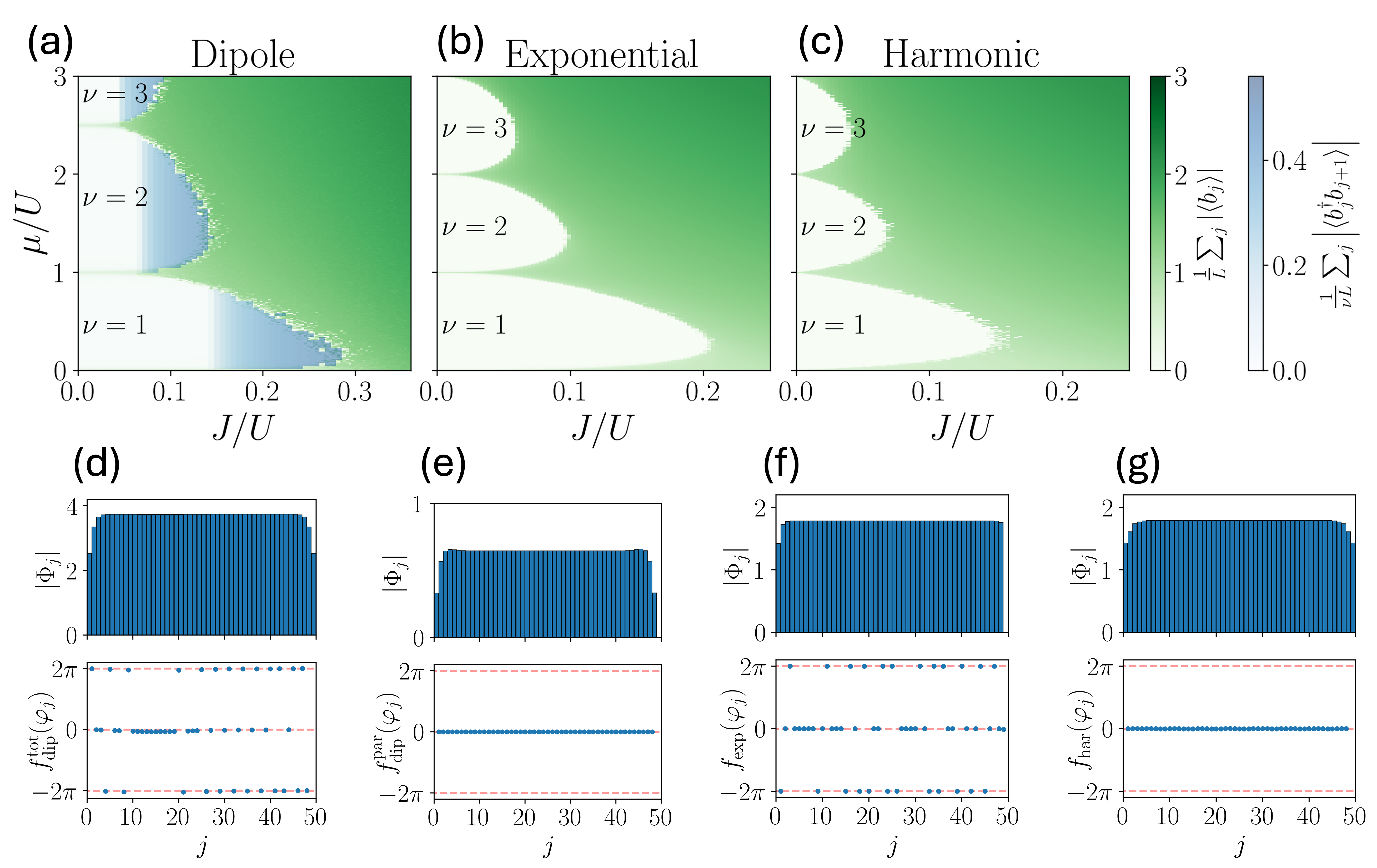}
\caption{(a)-(c): Mean-field phase diagram of deformed Bose-Hubbard models in \cref{table: various models} with (a) dipole ($U'=U/2$), (b) exponential ($U'=0$), and (c) harmonic ($U'=0$) TCMSs, computed for $L=50$ sites with open boundary condition and a maximum on-site boson number cutoff $N_{\text{max}}=20$ (see numerical methods in the SM \cite{SM}). (d)-(g): The order parameters $\Phi_j = \lket{\GS} \orderP{j} \rket{\GS} = |\Phi_j| e^{i \goldstone_j}$ of the SSB phases with respect to spatial position $j$. Panels (d),(f) and (g) show $\Phi_j=\langle \hb_j \rangle$ at $(\mu/U, J/U)=(1.5, 0.3)$ for the dipole model, exponential model and harmonic model, respectively. Panel (e) shows $\Phi_j=\langle \hb_j^{\dagger} \hb_{j+1} \rangle$ at $(\mu/U, J/U)=(0.3, 0.2)$ for the dipole model. In each panel, the upper plot shows the order parameter magnitude $|\Phi_j|$ is spatially uniform in the bulk, while the lower plot shows the SSB order parameters have the following phases angles be zero: (d): $f_{\text{dip}}^{\text{tot}}(\goldstone_j) = \goldstone_{j-1} - 2 \goldstone_{j} + \goldstone_{j+1}$. (e): $f_{\text{dip}}^{\text{par}}(\goldstone_j) = \goldstone_{j} - \goldstone_{j-1}$. (f): $f_{\text{exp}}(\goldstone_j) = \goldstone_{j-1} - 2 \goldstone_{j}$. (g): $f_{\text{har}}(\goldstone_j) = \goldstone_{j-1} - \goldstone_{j} + \goldstone_{j+1}$. 
}
\label{fig: phase diagram}
\end{figure}

\emph{Goldstone action}. The translationally symmetric models with a TCMS can generically exhibit spontaneous symmetry breaking (SSB) phases (above the Mermin-Wagner critical dimension \cite{Mermin_1966,Hohenberg_1967}) breaking the charges $\hQset$. \cref{fig: phase diagram} shows the Gutzwiller mean field phase diagram of the dipole, exponential and harmonic models in \cref{table: various models}, respectively, all of which enter SSB phases at sufficiently large $J/U$ (similar to the standard Bose-Hubbard model \cite{Bose_Hubbard_Fisher}). Interestingly, all the SSB phases have spatially uniform magnitudes of order parameters and particle density in the bulk (\cref{fig: phase diagram} (d-g), implying an unbroken symmetry involving translation. Particularly, the dipole symmetry model has an SSB phase (blue) breaking only the dipole charge, and another SSB phase (green) breaking both the monopole and dipole charges \cite{Lake_2022}. Some of these SSB phases are found to have unconventional Goldstone mode properties \cite{Stahl_2022,hu2024bosonicquantumbreakdownhubbard,hu2025quantumbreakdowncondensatedisorderfree}.

We are thus motivated to consider generic SSB phases of TCMS with unbroken translation symmetry $\hP_\mu$ (up to ground-state-dependent conjugation by $\hQset$), to derive their low-energy Goldstone actions irrespective of the microscopic models. This amounts assuming the existence of one translationally symmetric SSB ground state, denoted as $|\Omega_{\boldsymbol{0}}\rangle$, which satisfies $\hP_\mu|\Omega_{\boldsymbol{0}}\rangle=p_\mu^{(0)}|\Omega_{\boldsymbol{0}}\rangle$ with some momentum eigenvalues $p_\mu^{(0)}$. 

Assume the SSB is characterized by a minimal set of order parameters being ground state expectations $\langle \orderP{a}(\boldx)\rangle$ ($1\le a\le \Lambda$), where $\orderP{a}(\boldx)$ are charged local operators obeying the translational invariance $\orderP{a}(\boldx)=e^{-i\hP_\mu x^\mu}\orderP{a}(\boldsymbol{0})e^{i\hP_\mu x^\mu}$. We can assume each $\orderP{a}(\boldx)$ occupies an irreducible representation of the Abelian modulated charges $\hQset$, which is one-dimensional, and thus 
\begin{equation}\label{eqn:order-charge}
[\hQset,\orderP{a}(\boldx)]=\qSet_a(\boldx)\orderP{a}(\boldx)\ ,\quad \qSet_a(\boldx)=e^{A_\mu x^\mu}\qSet_a(\boldsymbol{0})\ ,
\end{equation}
where $\qSet_a(\boldx)=(q_{a,1}(\boldx),\cdots q_{a,N}(\boldx))^T$ is the modulated charge of $\orderP{a}(\boldx)$ in vector form. The latter equation in \cref{eqn:order-charge} is derived by noting that $\qSet_a(\boldx)\orderP{a}(\boldx)=[\hQset,e^{-i\hP_\mu x^\mu}\orderP{a}(\boldsymbol{0})e^{i\hP_\mu x^\mu}]=e^{A_\mu x^\mu}\qSet_a(\boldsymbol{0})\orderP{a}(\boldx)$. To form a complete set of order parameters, for each charge $\hQ^{(m)}$, one requires some index $a$ such that $q_{a,m}(\boldx)\neq 0$, so that the charge $\hQ^{(m)}$ is broken (unbroken) if $\langle \orderP{a}(\boldx)\rangle\neq 0$ ($\langle \orderP{a}(\boldx)\rangle\equiv 0$).

We now assume the SSB phase has $N_\text{ssb}\le N$ spontaneously broken charges denoted as $\overline{\hQset}\subseteq \hQset$ in vector form, and $N-N_\text{ssb}$ unbroken charges denoted as $\hQset'\subseteq \hQset$ in vector form ($\overline{\hQset}\cup\hQset'=\hQset$). The allowed SSB phases must have the unbroken charges $\hQset'$ form a closed subalgebra under translation:
\begin{equation}\label{eqn:unbroken-Q-commutator}
[\hP_\mu,\, \hQset']=-iA_\mu'\hQset'\ ,
\end{equation}
so that they can form a stabilizer subgroup, where $A_\mu'$ can be read off from the matrix $A_\mu$. Accordingly, translation and the broken charges $\overline{\hQset}$ have commutators 
\begin{equation}\label{eqn:broken-Q-commutator}
[\hP_\mu,\, \overline{\hQset}]
= -i \overline{A}_{\mu}\, \overline{\hQset}
-iA_\mu''\hQset'\ ,
\end{equation}
where $\overline{A}_{\mu}$ and $A_\mu''$ can be read off from the matrix $A_\mu$. (If all charges $\hQset$ are broken, one has $\overline{A}_{\mu}=A_\mu$.) From the reference SSB ground state $\rket{\GS_{\boldsymbol{0}}}$, all the SSB ground states can be generated by a symmetry transformation as $\rket{\GS_{\boldsymbol{\xi}}} = e^{i \boldsymbol{\xi}^{T} \overline{\hQset}} \rket{\GS_{\boldsymbol{0}}}$, which form a coset parameterized by a $N_\text{ssb}$-component vector $\boldsymbol{\xi}$. By \cref{eqn:broken-Q-commutator}, each SSB ground state $\rket{\GS_{\boldsymbol{\xi}}}$ is an eigenstate of $\hP_{\mu}-\boldsymbol{\xi}^{T} \overline{A}_{\mu} \overline{\hQset}$, which indicates an unbroken mixed translational symmetry.
From \cref{eqn:order-charge}, the order parameters of SSB ground states $\rket{\GS_{\boldsymbol{\xi}}}$ and $\rket{\GS_{\boldsymbol{0}}}$ are related by
\begin{equation}\label{eqn:order-trans}
\lket{\GS_{\boldsymbol{\xi}}}\orderP{a}(\boldx)\rket{\GS_{\boldsymbol{\xi}}} = 
e^{-i \boldsymbol{\xi}^{T} e^{\overline{A}_{\mu} x^{\mu}} \overline{\qSet}_a(\boldsymbol{0})}
\lket{\GS_{\boldsymbol{0}}}\orderP{a}(\boldx)\rket{\GS_{\boldsymbol{0}}}\ ,
\end{equation}
where $\overline{\qSet}_a(\boldsymbol{0})$ is $\qSet_a(\boldsymbol{0})$ restricted in the basis of broken charges.

\begin{table}[t]
\centering
\caption{The example Goldstone Lagrangian density $\mathcal{L}=\mathcal{L}_0+\mathcal{L}_1$ for three representative cases of 1D TCMS, where $\mathcal{L}_0=\frac{1}{2}(\dot{\goldstoneSet}^{T} \dot{\goldstoneSet}-\sum_{\mu=2}^d \partial_\mu\goldstoneSet^{T}\partial_\mu\goldstoneSet)$ in $d$-dimensional space.}
\label{table: goldstone}
\begin{tabular}{|c|c|c|}
\hline
TCMS  & broken $\overline{\hQset}$ & Goldstone Lagrangian $\mathcal{L}_1$ \\
\hline
\hline
\multirow{2}{*}{dipole} & $\hQmul{\text{d}},\hQmul{\text{m}}$ & $-\frac{v_\text{d}^2}{2}(\partial_1\varphi_\text{d})^2-\frac{v_\text{m}^2}{2}(\partial_1\varphi_\text{m}-\varphi_\text{d})^2$ \\
\cline{2-3}
 & $\hQmul{\text{d}}$ & $-\frac{v_\text{d}^2}{2}(\partial_1\varphi_\text{d})^2$ \\
\hline
exponential & $\hQexp$ & $-\frac{v_\text{e}^2}{2}(\partial_1\varphi_\text{e}+\alpha \varphi_\text{e})^2$ \\
\hline
harmonic & $\hQsin,\hQcos$ & $-\frac{v_\text{h}^2}{2}\left[(\partial_1\varphi_\text{s}+\beta \varphi_\text{c})^2+(\partial_1\varphi_\text{c}-\beta \varphi_\text{s})^2\right]$ \\
\hline
\end{tabular}
\end{table}

The translation symmetry of state $\rket{\GS_{\boldsymbol{0}}}$ ensures that $\lket{\GS_{\boldsymbol{0}}}\orderP{a}(\boldx)\rket{\GS_{\boldsymbol{0}}}=\Phi_{a,0}$ is independent of $x$. To derive the Goldstone action, we can promote the order parameter into $\langle\orderP{a}(\boldx,t)\rangle=\Phi_{a,0} e^{-i\goldstoneSet^T(\boldx,t)\overline{\qSet}_a(\boldsymbol{0})}$, where $\goldstoneSet(\boldx,t)$ is an $N_\text{ssb}$-component real Goldstone field. Under a symmetry transformation $e^{i \boldsymbol{\xi}^{T} \overline{\hQset}}$ in the ground state coset, $\goldstoneSet(\boldx,t)$ transforms according to \cref{eqn:order-trans} as:
\begin{equation}\label{eqn:goldstone-trans}
\goldstoneSet (\boldx,t) \rightarrow \goldstoneSet (\boldx,t) + e^{\overline{A}_{\mu}^T x^{\mu}} \boldsymbol{\xi}\ .
\end{equation}
The Goldstone action must be invariant under the transformation of \cref{eqn:goldstone-trans}. This requires it to be a function of the time derivative $\dot{\goldstoneSet}=\partial_t\goldstoneSet$ and spatial derivatives $\boldsymbol{\Omega}_\mu=(\partial_\mu - \overline{A}_{\mu}^T) \goldstoneSet$ (see SM \cite{SM} for a coset construction proof \cite{Coleman:1969sm,Callan:1969sn} that gives the same result). To the quadratic order, this gives a generic Lagrangian density for the Goldstone action:
\begin{equation}\label{eqn:Goldstone-L}
    \mathcal{L} = 
    \frac{1}{2} \dot{\goldstoneSet}^{T} M \dot{\goldstoneSet} 
    - \frac{1}{2} \boldsymbol{\Omega}_{\mu}^{T} W^{\mu\nu} \boldsymbol{\Omega}_{\nu}\ ,
\end{equation}
where $M=M^T$ and $W^{\mu\nu}=(W^{\nu\mu})^T$ ($\mu,\nu=1,\cdots,d$) are real $N_\text{ssb}\times N_\text{ssb}$ matrices such that $M$ is positive definite, and $\boldsymbol{\Omega}_{\mu}^{T} W^{\mu\nu} \boldsymbol{\Omega}_{\nu}$ is positive for nonzero $\boldsymbol{\Omega}_\mu$. Here we assumed the leading term in $\dot{\goldstoneSet}$ is quadratic, which is for instance ensured by an unbroken time-reversal symmetry.

\cref{eqn:Goldstone-L} gives an equation of motion for the Fourier Goldstone mode $\widetilde{\goldstoneSet}_{\boldsymbol{k},\omega}$ at energy $\omega$ and momentum $\boldsymbol{k}$:
\begin{equation}
\left[-\omega^2 M + (k_\mu-i\overline{A}_\mu) W^{\mu\nu}(k_\nu +i \overline{A}_\nu^T)\right]\widetilde{\goldstoneSet}_{\boldsymbol{k},\omega}=0\ ,
\end{equation}
from which the Goldstone mode dispersions can be derived. Particularly, an $\omega=0$ gapless mode occurs at momenta $\boldsymbol{k}$ where all matrices $k_\mu+i \overline{A}_\mu^T$ have a common eigenvector with zero-eigenvalue. A spectral argument on how this criterion modifies the Goldstone theorem for TCMS is given in the End Matter. We now examine three representative cases with 1D TCMS in \cref{table: goldstone}.

\begin{figure}[t]
\centering
\includegraphics[width=90mm]{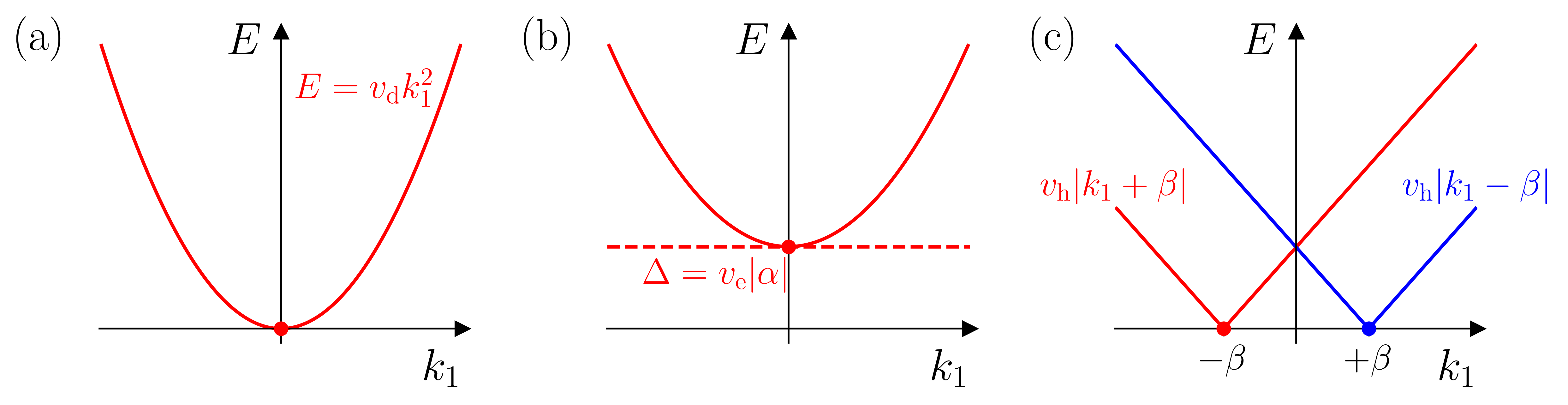}
\caption{Schematic dispersions of the Nambu-Goldstone modes arising from spontaneous breaking of the three classes of modulated symmetries, assuming $k_\perp=0$. (a)~Multipole: a single gapless mode with quadratic dispersion $E = v_\text{d} k_1^2$, consistent with the conventional Goldstone theorem. (b)~Exponential: an intrinsically gapped mode with dispersion $E = v_\text{e} \sqrt{\alpha^2 + k_1^2}$ and a gap $\Delta = v_\text{e} |\alpha|$ set by the exponential weight. (c)~Harmonic: two gapless branches $E = v_\text{h} |k_1 \pm \beta|$ with minima shifted to finite momenta $k_1 = \pm\beta$.}
\label{fig: dispersion}
\end{figure}

In a dipole symmetry model with $A_1=\JordanReal_2(0)$ and all $A_{\mu\ge2}=0$, one has $\hQset=(\hQmul{\text{d}}, \hQmul{\text{m}})^{T}$, where $\hQmul{\text{d}}$ is the dipole charge, and $\hQmul{\text{m}}$ is the monopole charge (\cref{table: various models}). There are two SSB phases satisfying \cref{eqn:unbroken-Q-commutator}: phase-I (green in \cref{fig: phase diagram}(a)) has both $\hQmul{\text{d}}$ and $\hQmul{\text{m}}$ broken, yielding a two-component Goldstone field $\goldstoneSet=(\varphi_\text{d},\varphi_\text{m})^T$. An example Goldstone Lagrangian is given in \cref{table: goldstone}. This yields a gapless mode and a gapped mode, which have anisotropic small $\boldsymbol{k}$ dispersions $\omega_-=\sqrt{v_\text{d}^2 k_1^4+k_\perp ^2}$ and $\omega_+=v_\text{m}+\frac{(v_\text{d}^2+v_\text{m}^2)k_1^2+k_\perp^2}{2v_\text{m}}$, respectively, with $k_\perp=\sqrt{\boldsymbol{k}^2-k_1^2}$ (see SM \cite{SM} for derivations).
Note that the gapless mode $\omega_-$ is quadratic in $k_1$ as shown in \cref{fig: dispersion}(a), in agreement with the finding of \cite{Stahl_2022}. Phase-II (blue in \cref{fig: phase diagram}(a)) has only $\hQmul{\text{d}}$ broken. The corresponding Lagrangian of Goldstone field $\varphi_\text{d}$ (\cref{table: goldstone}) yields a gapless mode with a conventional linear dispersion $\omega=\sqrt{v_\text{d}^2 k_1^2+k_\perp ^2}$.

In an exponential symmetry model with $A_1=\JordanReal_1(\alpha)$ ($\alpha\neq0$) and $A_{\mu\ge2}=0$, one has an exponential charge $\hQexp$ (e.g. in \cref{table: various models}). In its SSB phase, an example action of the Goldstone field $\varphi_\text{e}$ is given in \cref{table: goldstone}, which gives a single gapped mode with dispersion $\omega=\sqrt{v_\text{e}^2 \alpha^2+v_\text{e}^2 k_1^2 +k_\perp^2}$ (\cref{fig: dispersion}(b)). Therefore, the SSB of a continuous exponential symmetry does not yield a gapless Goldstone mode, which implies a robust SSB even in 1D space. This agrees with the finding in the quantum breakdown model with an exponential symmetry \cite{hu2024bosonicquantumbreakdownhubbard}, while our result here is model independent.

In a harmonic symmetry model with $A_1=\JordanComplex_1(0,\beta)$ ($\beta\neq0$) and $A_{\mu\ge2}=0$, there are two symmetry charges denoted as $\hQset=(\hQsin,\hQcos)^{T}$. \cref{eqn:unbroken-Q-commutator} requires both charges to be broken simultaneously, leading to only one SSB phase. An example action of their Goldstone field $\goldstoneSet=(\varphi_\text{s},\varphi_\text{c})$ takes the form in \cref{table: goldstone}, which gives rise to two gapless Goldstone modes with dispersions $\omega_\pm=\sqrt{v_\text{h}^2(k_1\pm\beta)^2+k_\perp^2}$ gapless at momenta $(k_1,k_\perp)=(\pm\beta,0)$ (see derivations in SM \cite{SM}), as shown in \cref{fig: dispersion}(c). This is consistent with the linear modes found in \cite{Sala_2024_exotic}, although they studied a 1D harmonic symmetry model with no SSB due to Mermin-Wagner theorem \cite{Mermin_1966,Hohenberg_1967}, and thus the linear modes are not Goldstone but Luttinger modes.

\emph{Discussion}. We have introduced the definition of continuous TCMSs which allow translationally symmetric Hamiltonians, and showed that an Abelian TCMS can only contain multipole, exponential and harmonic components. Specifically, we classified 1D TCMSs by real Jordan normal forms. We further derived the generic Goldstone action for SSB phases of TCMSs, by which we showed that a broken multipole symmetry yields gapless Goldstone modes with higher-order dispersions, a broken harmonic symmetry yields gapless Goldstone modes at finite momenta, while a broken exponential symmetry gives no gapless Goldstone modes. An alternative spectral derivation of this modification to Goldstone theorem is presented in the End Matter. Future open questions include the classification of higher dimensional TCMSs and non-Abelian TCMSs, and simple models with such symmetries and possible exotic ground state properties therein. It would also be interesting to explore deeper understandings of matrices $A_\mu$ as an imaginary connection, and generalization of TCMSs in curved spacetimes.

\begin{acknowledgments}
\emph{Acknowledgments}. We thank Yumin Hu, Yu-Ping Wang, Junjia Zhang, Wucheng Zhang for helpful discussions. This work is supported by the National Science Foundation under award DMR-2141966, and the National Science Foundation through Princeton University’s Materials Research Science and Engineering Center DMR-2011750. 
\end{acknowledgments}

\newpage

\begin{center}
\textbf{END MATTER}
\end{center}

\emph{Constraints on the non-Abelian TCMS structure constants.}
The structure constants $B_{\ell}^{mn}$ (which are nonzero for non-Abelian TCMSs), defined through the commutation relation $[\hQ^{(m)},\hQ^{(n)}]=i B^{mn}_\ell \hQ^{(\ell)}$, are not arbitrary, but must satisfy several consistency conditions. First, antisymmetry of the commutator implies $B^{mn}_\ell = -B^{nm}_\ell$. In addition, the Jacobi identity involving $\hP_\mu$, $\hQ^{(m)}$, and $\hQ^{(n)}$ gives
\begin{equation}
\begin{aligned}
    B^{mn}_{\ell} (A_\mu)^{\ell}_{k}
    ,
    &=
    B^{\ell n}_{k} (A_\mu)^{m}_{\ell}
    +
    B^{m \ell}_{k} (A_\mu)^{n}_{\ell}
.\end{aligned}
\end{equation}
Furthermore, applying the Jacobi identity to $\hQ^{(k)}$, $\hQ^{(m)}$, and $\hQ^{(n)}$ yields
\begin{equation}
\begin{aligned}
    B^{mn}_{\ell} B^{k \ell}_{p}
    &=
    B^{km}_{\ell} B^{\ell n}_{p}
    +
    B^{kn}_{\ell} B^{m \ell}_{p}
.\end{aligned}
\end{equation}

\emph{The Modified Goldstone Theorem for TCMS.} 
Here, we derive the modification of the Goldstone theorem for continuous Abelian TCMS from the commutation relation between the TCMS charges and the momentum operator.

Consider a spontaneously broken symmetry charge of the form
$\hQ(t) = \int \modulatingFunctionIndex{}{\boldx} \hj_{0}(\boldx, t) d^d \boldx$, where $\hj_{0}(\boldx, t)$ is a local Hermitian operator in Heisenberg picture satisfying commutation relations with the translation operators $\hP_\mu$ ($1\le \mu\le d$) and the Hamiltonian $H$:
\begin{equation}\label{eq:comm-P-H-j0}
[\hP_\mu, \hj_{0}(\boldx, t)]=i \partial_\mu \hj_{0}(\boldx, t),\ [H, \hj_{0}(\boldx, t)]=-i\partial_t \hj_{0}(\boldx, t). 
\end{equation}
Assume $\hat{\boldsymbol{j}}(\boldx, t)$ is the charge current density satisfying the continuity equation $\modulatingFunctionIndex{}{\boldx} \partial_t \hj_{0}(\boldx, t) + \nabla \cdot \hat{\boldsymbol{j}}(\boldx, t) = 0$, such that the charge satisfies $d\hQ(t)/dt=0$. For conventional symmetries, $\modulatingFunctionIndex{}{\boldx}=1$, while for TCMS, $\modulatingFunctionIndex{}{\boldx}$ is $\boldx$-dependent. By \cref{eq:comm-P-H-j0}, the charge density $\hj_{0}(\boldx, t)$ can be expressed as a spacetime translation of $\hj_{0}(\bold{0}, 0)$: 
\begin{equation}
    \hj_{0}(\boldx, t) = 
    e^{i \hH t-i \hP_\mu x^\mu}
    \hj_{0}(\bold{0}, 0)
    e^{-i \hH t+i \hP_\mu x^\mu}
\end{equation}
Assume the symmetry generated by $\hQ$ is spontaneously broken by a ground state $\rket{\GS}$. This means that there exists a charged local operator (playing the role of the order parameter) $\orderP{}(\boldx, t)$ such that $\lket{\GS}[\hQ, \orderP{}(\boldx, t)]\rket{\GS} \neq 0$. Without loss of generality, we assume the ground state $|\Omega\rangle$ has zero energy and zero momentum. Consider the spectral decomposition by inserting a complete set of energy-momentum eigenbasis $\{ \rket{n, \boldsymbol{k}} \}$ satisfying $H\rket{n, \boldsymbol{k}}=E_n(\boldsymbol{k})\rket{n, \boldsymbol{k}}$ and $\hP_\mu\rket{n, \boldsymbol{k}}=k_\mu\rket{n, \boldsymbol{k}}$, with energy $E_n(\boldsymbol{k})\ge0$ and momentum $\boldsymbol{k}=(k_1,\cdots,k_d)$. It suffices to consider the charged local operator (order parameter) at time zero, and using the fact that the conserved charge $\hQ=\hQ(0)=\hQ(t)$ is time independent, which gives:
\begin{widetext}
\begin{equation}
\begin{aligned}
\label{eqn: Goldstone theorem}
    & \lket{\GS}[\hQ, \orderP{}(\boldx, 0)]\rket{\GS}=\lket{\GS}[\hQ(t), \orderP{}(\boldx, 0)]\rket{\GS} 
    \\
    =&
    \int d^{d} \boldx' \modulatingFunctionIndex{}{\boldx'} \sum_n \int\frac{d^{d} \boldsymbol{k}}{(2\pi)^d} \left[\lket{\GS}\hj_{0}(\boldx', t)\rket{n, \boldsymbol{k}} \lket{n, \boldsymbol{k}} \orderP{}(\boldx, 0)\rket{\GS} 
    - \lket{\GS} \orderP{}(\boldx, 0) \rket{n, \boldsymbol{k}} \lket{n, \boldsymbol{k}} \hj_{0}(\boldx', t) \rket{\GS} \right]
    \\
    =&
    \int d^{d} \boldx' \modulatingFunctionIndex{}{\boldx'} \sum_n \int\frac{d^{d} \boldsymbol{k}}{(2\pi)^d} 
    \left[ e^{i \boldsymbol{k} \cdot \boldx'-i E_n(\boldsymbol{k}) t} \lket{\GS}\hj_{0}(0, 0)\rket{n, \boldsymbol{k}} \lket{n, \boldsymbol{k}} \orderP{}(\boldx, 0)\rket{\GS} 
    -  
    \text{h.c.}\right]
    \\
    =& \sum_n \int d^{d} \boldsymbol{k} \left[
    \Delta(\boldsymbol{k}) e^{-i E_n(\boldsymbol{k}) t} \lket{\GS}\hj_{0}(0, 0)\rket{n, \boldsymbol{k}} \lket{n, \boldsymbol{k}} \orderP{}(\boldx, 0)\rket{\GS} 
    - 
    \Delta(-\boldsymbol{k}) e^{i E_n(\boldsymbol{k}) t} \lket{\GS} \orderP{}(\boldx, 0) \rket{n, \boldsymbol{k}} \lket{n, \boldsymbol{k}} \hj_{0}(0, 0) \rket{\GS}
    \right],
\end{aligned}
\end{equation}
\end{widetext}
where $\text{h.c.}$ stands for Hermitian conjugate, and
\begin{equation}\label{eq:Delta-Goldstone}
    \Delta(\boldsymbol{k}) = \frac{1}{(2\pi)^d} \int d^{d} \boldx e^{i \boldsymbol{k} \cdot \boldx} f(\boldx)\ .
\end{equation}
For conventional symmetries with $\modulatingFunctionIndex{}{\boldx}=1$, one has $\Delta(\boldsymbol{k}) = \delta(\boldsymbol{k})$, and this implies that only eigenstates in the $\boldsymbol{k}\rightarrow\bold{0}$ limit contribute to the integral. It is easy to see the left hand side (LHS) of \cref{eqn: Goldstone theorem} is independent of time $t$, while the right hand side (RHS) of \cref{eqn: Goldstone theorem} has $t$-dependent coefficients $e^{\pm iE_n(\boldsymbol{k})t}$ if $E_n(\boldsymbol{k})>0$, so the terms with $E_n(\boldsymbol{k})>0$ must vanish. Since the LHS is nonzero in the SSB state, there must exist gapless modes contributing to the RHS, satisfying $\min_n\lim_{\boldsymbol{k} \rightarrow \bold{0}} E_n(\boldsymbol{k}) = 0$.  This proves the conventional Goldstone theorem.

In the case of TCMS with $\modulatingFunctionIndex{}{\boldx}$ depending on $\boldx$, the function $\Delta(\boldsymbol{k})$ in \cref{eq:Delta-Goldstone} will change, and the Goldstone theorem will be modified accordingly. As examples, we examine the following simple cases of 1D TCMS, which have $\modulatingFunctionIndex{}{\boldx}=\modulatingFunctionIndex{}{x^1}$ only depending on the coordinate $x^1$ of the first dimension.

(i) In a model with multipole symmetry, the $p$-th multipole symmetry charge $\hQ_p$ has a unit-of-charge function $\modulatingFunctionIndex{}{\boldx}=\modulatingFunctionIndex{}{x^1}=\frac{(x^1)^p}{p!}$ (see the paragraph below \cref{eqn:Jcomplex}). If charge $\hQ_p$ is broken, the corresponding function in \cref{eq:Delta-Goldstone} is $\Delta(\boldsymbol{k})=\frac{1}{p!}(-i\partial_{k_1})^p\delta(\boldsymbol{k})$. Such a function $\Delta(\boldsymbol{k})$ is nonzero only when $\boldsymbol{k}\rightarrow\bold{0}$, so \cref{eqn: Goldstone theorem} implies the existence of gapless modes satisfying $\min_n\lim_{\boldsymbol{k} \rightarrow \bold{0}} E_n(\boldsymbol{k}) = 0$, in agreement with our conclusions in the main text.

(ii) In a model with an exponential symmetry, the exponential charge $\hQexp$ has a unit-of-charge function $\modulatingFunctionIndex{}{\boldx}=\modulatingFunctionIndex{}{x^1}=e^{\alpha x^1}$, where $\alpha\neq0$. For any real momentum $\boldsymbol{k}$, \cref{eq:Delta-Goldstone} gives a diverging function $\Delta(\boldsymbol{k})\rightarrow\infty$ in an infinite space, and thus in \cref{eqn: Goldstone theorem}, as long as the LHS is finite, the matrix elements on the RHS can be zero. Therefore, no gapless modes are guaranteed to exist in the bulk, in agreement with our conclusion in the main text. If the space has an open boundary, in which case edge modes can have complex valued momenta $\boldsymbol{k}$, \cref{eq:Delta-Goldstone} gives a function $\Delta(\boldsymbol{k})=\delta(\boldsymbol{k}-i\alpha \hat{\boldsymbol{e}}_1)$ in the sense of analytical continuation (note that it no longer diverges when the space has a finite boundary), where $\hat{\boldsymbol{e}}_1$ is the unit vector in the first dimension. This implies the existence of gapless edge modes at imaginary momentum $\boldsymbol{k}=i\alpha\hat{\boldsymbol{e}}_1$ (i.e., decay length $1/\alpha$ in the $x^1$ direction), $\min_n\lim_{\boldsymbol{k}\rightarrow i\alpha \hat{\boldsymbol{e}}_1} E_n(\boldsymbol{k}) = 0$ in agreement with the findings in \cite{hu2024bosonicquantumbreakdownhubbard,hu2025quantumbreakdowncondensatedisorderfree}.

(iii) In a model with a harmonic symmetry, the two charges $\hQsin,\hQcos$ have unit-of-charge functions $f_\text{s}(\boldx)=f_\text{s}(x^1)=\sin{\beta x^1}$ and $f_\text{c}(\boldx)=f_\text{c}(x^1)=\cos{\beta x^1}$, respectively, which are always simultaneously broken. \cref{eq:Delta-Goldstone} gives functions $\Delta_\text{s}(\boldsymbol{k})=\frac{i}{2}\left[\delta(\boldsymbol{k}-\beta \hat{\boldsymbol{e}}_1)-\delta(\boldsymbol{k}+\beta \hat{\boldsymbol{e}}_1)\right]$, and $\Delta_\text{c}(\boldsymbol{k})=\frac{1}{2}\left[\delta(\boldsymbol{k}-\beta \hat{\boldsymbol{e}}_1)+\delta(\boldsymbol{k}+\beta \hat{\boldsymbol{e}}_1)\right]$, respectively. With \cref{eqn: Goldstone theorem}, both broken charges imply the existence of gapless modes at finite momenta $\boldsymbol{k}=\pm \beta \hat{\boldsymbol{e}}_1$, i.e., $\min_n\lim_{\boldsymbol{k}\rightarrow \pm \beta \hat{\boldsymbol{e}}_1} E_n(\boldsymbol{k}) = 0$, in agreement with our conclusion in the main text.

\bibliography{mybib.bib}

\clearpage
\onecolumngrid

\setcounter{equation}{0}
\setcounter{figure}{0}
\setcounter{table}{0}
\setcounter{page}{1}
\makeatletter
\renewcommand{\theequation}{S\arabic{equation}}
\renewcommand{\thefigure}{S\arabic{figure}}
\renewcommand{\thetable}{S\arabic{table}}

\begin{center}
{\bf \large Supplemental Material}
\end{center}

\section{The Baker-Campbell-Hausdorff (BCH) formula}

The Baker-Campbell-Hausdorff (BCH) formula is useful in deriving the equations in this paper:

\begin{equation}
\begin{aligned}
e^A B e^{-A} &= B + [A, B] + \frac{1}{2!}[A, [A, B]] + \frac{1}{3!}[A, [A, [A, B]]] + \cdots
\\
e^A e^{B} e^{-A} &= \exp\left( B + [A, B] + \frac{1}{2!}[A, [A, B]] + \frac{1}{3!}[A, [A, [A, B]]] + \cdots \right)
\end{aligned}
\end{equation}

\section{Categorizing 1D translationally covariant modulated symmetries by solving differential equations}
\label{sectionSM: Categorizing modulated symmetries by solving differential equations}
In the main text, we classified the 1D translationally covariant modulated symmetries (TCMSs) in one spatial dimension by block-diagonalizing the structure constant matrices into their Jordan forms. Here we show that the same classification can be obtained by solving the corresponding differential equations. Consider charges $\hQ^{(m)}$ defined as the spatial integral of the corresponding density $\hj_0^{(m)}$ with a spatially modulating weighting function: $\hQ^{(m)}(t)=\int \modulatingFunctionIndex{m}{x^1}\hj_0^{(m)}(\boldx, t) d^d \boldx$. Using the commutation relation $[\hP_1,\hj_0^{(m)}(\boldx, t)]=i\partial_1 \hj_0^{(m)}(\boldx, t)$, one finds $[\hP_1, \hQ^{(m)}(t)]=-i \int \partial_1 \modulatingFunctionIndex{m}{x^1} \hj_{0}^{(m)}(\boldx, t) d^d \boldx$, where boundary terms are assumed to vanish. In general, the local operators $\hj_0^{(m)}(\boldx, t)$ differ for different charges, so each charge transforms irreducibly under translation. However, if one considers a finite set of $N$ charges $\hQset = (\hQ^{(1)},\dots,\hQ^{(N)})^{T}$ that share the same local density $\hat j_0^{(1)}=\cdots=\hat j_0^{(N)}$, they may mix under $\hP_1$. Suppose the commutation relation takes the form $[\hP_1,\, \hQset] = -i A_{1}\, \hQset$, where $A_1$ is the $N\times N$ structure constant matrix in the $x^1$-direction. By the Cayley–Hamilton theorem, there exists a set of real coefficients $\{ c_{k} \}_{k=0}^{N}$, such that $\sum_{k=0}^{N} c_k A_{1}^k = 0$. This relation can be translated into a linear differential constraint on $\modulatingFunctionIndex{m}{x^1}$:
\begin{equation}
    0 
    = \sum_{k=0}^{N} c_k A_1^k \hQset
    = \sum_{k=0}^{N} i^{k} c_k \underbrace{[\hP_1, \dots,[\hP_1}_{k}, \hQset]\dots]
    \Rightarrow
    \sum_{k=0}^{N} c_k \left(\frac{d}{dx^{1}}\right)^k \modulatingFunctionIndex{m}{x^1} = 0
\end{equation}
The roots of the characteristic polynomial $\sum_{k=0}^{N} c_k \lambda^k = 0$ then determine the solution to the differential equation, and are summarized in \cref{table: categories of modulated symmetries from differential eq}. 

\begin{table}[h]
\centering
\caption{Table for various TCMSs. The first row corresponds to a conventional symmetry rather than a modulated symmetry, since it has no spatial dependence. Rows 2-4 represent the simplest families of TCMSs, with no mixing between multipole, exponential, and the harmonic components.}
\label{table: categories of modulated symmetries from differential eq}
\begin{tabular}{|l|c|c|c|c|}
\hline
 Solution & Number & Multi- & Exp- & Harmonic \\
\hline
\hline
non-degenerate $\lambda=0$ & 1 & \xmark & \xmark & \xmark \\
\hline
$p$-fold degenerate $\lambda=0$ & $p$ & \cmark & \xmark & \xmark \\
\hline
non-degenerate $\lambda \in \mathbb{R} \backslash\{0\}$ & 1 & \xmark & \cmark & \xmark \\
\hline
non-degenerate $\lambda, \lambda^{*} \in i \mathbb{R}\backslash\{0\}$ & 2 & \xmark & \xmark & \cmark \\
\hline
$p$-fold degenerate $\lambda \in \mathbb{R} \backslash\{0\}$ & $p$ & \cmark & \cmark & \xmark \\
\hline
non-degenerate $\lambda \in \mathbb{C}\backslash(\mathbb{R}\cup i\mathbb{R})$ & 2 & \xmark & \cmark & \cmark \\
\hline
$l$-fold degenerate $\lambda, \lambda^{*} \in i \mathbb{R}\backslash\{0\}$ & $2l$ & \cmark & \xmark & \cmark \\
\hline
$l$-fold degenerate $\lambda \in \mathbb{C}\backslash(\mathbb{R}\cup i\mathbb{R})$ & $2l$ & \cmark & \cmark & \cmark \\
\hline
\end{tabular}
\end{table}

\section{Mean field calculation of the phase diagram}

In this section, we describe the mean-field method used to obtain the phase diagrams of the Bose-Hubbard models considered in the main text. As discussed in the main text, spontaneous breaking of a modulated symmetry generated by $\hQ$ can be detected by an order parameter $\orderP{}(\boldx)$ that is charged under $\hQ$:
\begin{equation}
[\hQset,\orderP{}(\boldx)]=\qSet(\boldx)\orderP{}(\boldx)
.\end{equation}
Using $[\hn_j, \hb_{j'}] = -\hb_j \delta_{j j'}$, we find that the single boson operator $\hb_j$ is charged under the conserved symmetries relevant to the models considered here:
\begin{equation}
    [\hQmul{\text{m}}, \hb_j ] = - \hb_j
    ,\quad
    [\hQmul{\text{d}}, \hb_j ] = -j \hb_j
    ,\quad
    [\hQmul{\text{e}}, \hb_j ] = -2^{-j} \hb_j
    ,\quad
    [\hQmul{\text{s}}, \hb_j ] = - \sin\left(\frac{j\pi}{3}\right) \hb_j
    ,\quad
    [\hQmul{\text{c}}, \hb_j ] = - \cos\left(\frac{j\pi}{3}\right) \hb_j
\end{equation}
Therefore, a nonzero expectation value $\langle\hb_j\rangle \neq 0$ signals spontaneous breaking of the corresponding conserved symmetries.

In addition, the dipolar Bose-Hubbard model can exhibit a phase in which the dipole symmetry is broken while the monopole symmetry remains unbroken. To detect this phase, we use the order parameter $\orderP{} = \hb_j^{\dagger} \hb_{j+1}$. This operator is neutral under $\hQmul{\text{m}}$ but charged under $\hQmul{\text{d}}$:
\begin{equation}
    [\hQmul{\text{d}}, \hb_j^{\dagger} \hb_{j+1} ] = -\hb_j^{\dagger} \hb_{j+1}
    ,\quad
    [\hQmul{\text{m}}, \hb_j^{\dagger} \hb_{j+1}] = 0.
\end{equation}
Consequently, a phase with $\langle \hb_j^{\dagger} \hb_{j+1} \rangle \neq 0$ and $\langle \hb_j \rangle = 0$ signals spontaneous breaking of the dipole symmetry while preserving the monopole symmetry.

\subsection{Broken to $U(1)$ condensation}
The main idea of this section follows Ref. \cite{hu2024bosonicquantumbreakdownhubbard}. The $U(1)$ condensate phase in the mean-field phase diagram shown in the main text is identified by parameterizing the ground state (GS) using the Gutzwiller wave function:
\begin{equation}
\rket{\GS} = \prod_{j=1}^{L} \left( \sum_{n=0}^{N_{\text{max}}} c_{j, n} \frac{(\hb_j^\dagger)^n}{\sqrt{n!}} \right)
\rket{0}
\end{equation}
where $\ket{0}$ denotes the vacuum state satisfying $\hb_j \ket{0} = 0$ for all lattice sites $j$, and $c_{j,n}$ are variational coefficients.

Under this ansatz, the local order parameter can be expressed as
\begin{equation}
    \langle\orderP{j}\rangle
    = \langle \hb_j \rangle 
    = \frac{\lket{\GS} \hb_j \rket{\GS}}{\lrket{\GS}{\GS}}
    = \frac{\sum_{n=0}^{N_{\text{max}}-1} c_{j, n}^{*} c_{j, n+1} \sqrt{n+1}}{\sum_{n=0}^{N_{\text{max}}} |c_{j, n}|^2} 
\end{equation}
Additionally, the density operator and the local pairing function are given by:
\begin{equation}
    \begin{aligned}
        \langle \hn_j \rangle
        &= \frac{\lket{\GS} \hb_j^\dagger \hb_j \rket{\GS}}{\lrket{\GS}{\GS}}
        = \frac{\sum_{n=0}^{N_{\text{max}}} n |c_{j, n}|^2}{\sum_{n=0}^{N_{\text{max}}} |c_{j, n}|^2} 
        \\
        \langle \hb_j^2 \rangle
        &= \frac{\lket{\GS} \hb_j \hb_j \rket{\GS}}{\lrket{\GS}{\GS}}
        = \frac{\sum_{n=0}^{N_{\text{max}}-2} c_{j, n}^{*} c_{j, n+2} \sqrt{(n+1)(n+2)}}{\sum_{n=0}^{N_{\text{max}}} |c_{j, n}|^2} 
    \end{aligned}
\end{equation}
Therefore, the energy density can be expressed as:
\begin{equation}
    \mathcal{E}_{\GS}(\{ c_{j, n} \}) = \frac{1}{L} \frac{\lket{\GS} H \rket{\GS}}{\lrket{\GS}{\GS}}
    = \frac{1}{L} \sum_{j=1}^{L} \frac{\sum_{n=0}^{N_{\text{max}}}\left[-\mu n + U n (n-1)/2\right] |c_{j, n}|^2}{ \sum_{n=0}^{N_{\text{max}}} |c_{j, n}|^2} + \mathcal{E}_{\GS}^{\text{int}}(\{ c_{j, n} \})
\end{equation}
where $\mathcal{E}_{\GS}^{\text{int}}(\{ c_{j, n} \})$ is different for each deformed Bose-Hubbard model:
\begin{equation}
    \begin{cases}
        \text{Dipolar: } 
        &\mathcal{E}_{\GS}^{\text{int}}(\{ c_{j, n} \}) = -\frac{J}{L}\sum_{j=2}^{L-1} 
        \left( 
        \lrangle{\hb_{j-1}} \lrangle{\hb_{j}^2}^{*} \lrangle{\hb_{j+1}} 
        + \lrangle{\hb_{j-1}}^{*} \lrangle{\hb_{j}^2} \lrangle{\hb_{j+1}}^{*} 
        \right)
        \\
        \text{Exponential: } 
        &\mathcal{E}_{\GS}^{\text{int}}(\{ c_{j, n} \}) = -\frac{J}{L}\sum_{j=1}^{L-1} \left( 
        \lrangle{\hb_{j}} \lrangle{\hb_{j+1}^2}^{*} 
        + \lrangle{\hb_{j}}^{*} \lrangle{\hb_{j+1}^2} 
        \right)
        \\
        \text{Harmonic: } 
        &\mathcal{E}_{\GS}^{\text{int}}(\{ c_{j, n} \}) = -\frac{J}{L}\sum_{j=2}^{L-1} 
        \left( 
        \lrangle{\hb_{j-1}} \lrangle{\hb_{j}}^{*} \lrangle{\hb_{j+1}} 
        + \lrangle{\hb_{j-1}}^{*} \lrangle{\hb_{j}} \lrangle{\hb_{j+1}}^{*} 
        \right)
    \end{cases}
\end{equation}
In practice, we minimize the variational energy with respect to the complex parameters $c_{j, n}$, and obtain the phase diagram from the optimized GS across parameter space.

\subsection{Broken to dipole condensation}

In the dipolar Bose-Hubbard model, the dipole-condensed phase occurs within each Mott insulating phase with a fixed particle number $\nu$ on every site. We therefore adopt the following variational ansatz:
\begin{equation}
    \rket{\GS} = 
    \left[\prod_{j=1}^{L-1} (\alpha_{j} + u_{j} \hb_{j}^{\dagger} \hb_{j+1} + v_{j} \hb_{j+1}^{\dagger} \hb_{j})\right]
    \times
    \left[ \prod_{j=1}^{L} ( \hb_{j}^{\dagger} )^{\nu} \right] \rket{0}
,\end{equation}
where $u_j$ and $v_j$ are complex variables and $\alpha_j$ are real. Since the factors in the first product generally do not commute, there is no universal closed-form expression for the variational energy or for local order parameters such as $\langle \hb_{j}^{\dagger} \hb_{j+1} \rangle$ at arbitrary system size $L$. Nevertheless, for any fixed $L$, these quantities can be evaluated explicitly. By minimizing the energy density $\mathcal{E}_{\GS} = \frac{1}{L} \frac{\lket{\GS} H \rket{\GS}}{\lrket{\GS}{\GS}}$ with respect to $(\alpha_j, u_j, v_j)$, we determine whether the dipole symmetry is spontaneously broken: a dipole-condensed phase is signaled by nonzero expectation values of the local dipole order parameters $\langle \hb_{j}^{\dagger} \hb_{j+1} \rangle$.

\section{Brief Overview of the Coset Construction}

The coset construction provides a systematic framework for deriving low-energy effective theories with spontaneously broken continuous symmetries. It was developed in the context of nonlinear realizations of symmetry, and is often referred to as the Callan-Coleman-Wess-Zumino (CCWZ) construction \cite{Coleman:1969sm,Callan:1969sn}. The main idea is that, when a symmetry group $G$ is spontaneously broken to a subgroup $H$, the low-energy Goldstone fields are coordinates on the coset space $G/H$. Symmetry then strongly constrains the allowed terms in the effective action. In particular, the coset construction gives a practical method for determining how Goldstone fields transform under $G$, constructing covariant building blocks from the Maurer-Cartan (MC) form $\gamma$, and writing down the most general effective action consistent with the symmetry.

We denote the unbroken generators by $\hT_A$ and the broken generators by $\hbroken_m$, so that
$\mathfrak g=\mathfrak h\oplus\mathfrak m$, with $\mathfrak h=\mathrm{span}\{\hT_A\}$ and $\mathfrak m=\mathrm{span}\{\hbroken_m\}$. 
We assume the coset is reductive, $[\mathfrak h,\mathfrak m]\subset\mathfrak m$.

\subsection{Without spacetime symmetry}

Let us first review the case in which no spacetime symmetry is involved.
Spontaneous symmetry breaking implies that the GSs are degenerate. 
We can choose one reference GS $|\Omega_0\rangle$ satisfying
\begin{equation}
\label{eqnSM: h GS = GS}
    h \ket{\GS_0} = \ket{\GS_0} ~, \quad \forall h \in H 
,\end{equation}
up to a phase factor. Note that for quantum states, we always define $|\psi\rangle=|\varphi\rangle$ as two states equal up to a phase factor. 
Acting on $\ket{\GS_0}$ with elements of $G$ generates the manifold of degenerate GSs:
\begin{equation}
    \ket{\GS_0} \xrightarrow{g} \ket{\GS_g} \equiv g\ket{\GS_0} ~.
\end{equation}
However, two group elements $g_1, g_2 \in G$ generate the same GS whenever they are connected by the right multiplication of the group element $h\in H$:
\begin{equation}
    \ket{\GS_{g_1}} = \ket{\GS_{g_2}}\quad \text{iff} \quad g_2 \in g_1 H.
\end{equation}
Therefore, the degenerate GSs are naturally labeled by the left cosets $gH \in G/H$, where $G/H \equiv \{gH\,|\,g\in G\}$. 

To make this explicit, one may choose an overcomplete representative to label the GSs
\begin{equation}
    \hat{\mathcal{U}}(\goldstone, \theta) = 
    \prod_m \exp\Big(i \goldstone^m  \hbroken_m\Big) 
    \prod_A \exp\Big(i \theta^A \hT_A\Big)
    \quad \Longrightarrow \quad
    \ket{\GS_{\goldstone, \theta}} = \hat{\mathcal{U}}(\goldstone, \theta) \ket{\GS_0}
.\end{equation}
Here the particular ordering of the exponentials in $\hat{\mathcal{U}}(\goldstone, \theta)$ is only a choice of coordinates on the group manifold. 
Other orderings give different parametrizations, related by field redefinitions, and do not change the underlying physical content.
Since the unbroken generators leave $\ket{\GS_0}$ invariant (\cref{eqnSM: h GS = GS}), the coordinates $\theta^A$ do not label distinct GSs. Equivalently, representatives related by right multiplication by an element of $H$ describe the same point (GS) in the coset:
\begin{equation}
    \hat{\mathcal{U}}(\goldstone, \theta) 
    \,\sim\,
    \hat{\mathcal{U}}(\goldstone, \theta) h \;,
    \quad h \in H.
\end{equation}
The fields $\theta^A$ are therefore redundant $H$-frame coordinates rather than physical Goldstone fields. Since the unbroken generators leave the reference state invariant, the physical vacuum manifold is the quotient $G/H$. Equivalently, $G$ is the principal $H$-bundle over the base $G/H$, and the coordinates $\theta^A$ parametrize the $H$-fiber rather than independent physical vacua. To describe the base locally, we choose a section of the bundle. In the exponential coordinates above, a convenient local section is $\theta^A=0$ whose image is 
\begin{equation}
    \hU(\goldstone) = \hat{\mathcal{U}}(\goldstone, \theta=0)
    = \prod_m \exp\Big(i \goldstone^m  \hbroken_m\Big) 
.\end{equation}

\begin{figure}[t]
    \scalebox{0.8}{
        \begin{tikzpicture}[
    thick,
    >=Stealth,
    fiber/.style={dashed, blue!50!black, thick},
    gaction/.style={->, thick, green!50!black, 
        postaction={decorate, decoration={markings, mark=at position 0.5 with {\arrow{Stealth}}}}
    },
    haction/.style={->, thick, orange!80!black},
    dot/.style={circle, fill, inner sep=2.5pt},
]

\fill[blue!5] (0,-0.3) ellipse (4.5cm and 0.8cm);
\draw[blue!60!black, thick] (0,-0.3) ellipse (4.5cm and 0.8cm);

\node[below, blue!60!black, font=\large] at (0,-1.5) {$G/H$ \;(vacuum manifold)};

\node[dot, blue!70!black] (xi_base) at (-2.2,-0.2) {};
\node[below=3pt, blue!70!black, font=\normalsize] at (xi_base) {$\goldstone$};

\node[dot, red!70!black] (xip_base) at (2.0,-0.2) {};
\node[below=3pt, red!70!black, font=\normalsize] at (xip_base) {$\goldstone'$};

\draw[fiber] (-2.2,-0.2) -- (-2.2,5.5);
\node[above, blue!50!black, font=\small] at (-2.2,5.5) {$\cong H$};

\draw[fiber, red!50!black] (2.0,-0.2) -- (2.0,5.5);
\node[above, red!50!black, font=\small] at (2.0,5.5) {$\cong H$};

\draw[black!80, thick, dashed, rounded corners=12pt] (-5.0,-1.3) rectangle (5.5,6.0);
\node[black!80, font=\large, anchor=north east] at (5.5,5.9) {$G$};

\node[dot, blue!70!black] (Uxi) at (-2.2,2.5) {};
\node[left=5pt, blue!70!black, font=\normalsize, align=right] at (-2.1,2.1) {$\hU(\goldstone)$\\[-2pt]};

\node[dot, red!70!black] (Uxip) at (2.0,2.5) {};
\node[right=5pt, red!70!black, font=\normalsize,align=left] at (1.9, 2.1) {$\hU(\goldstone')$\\[-2pt]};

\node[dot, orange!80!black] (gUxi) at (2.0,4.2) {};
\node[right=5pt, orange!80!black, font=\normalsize, align=left] at (gUxi) {$g\,\hU(\goldstone)$\\[-2pt]};

\draw[black!80, thin, dashed] 
    (-4.0,2.8) .. controls (-3.2,2.4) and (-2.8,2.5) .. (Uxi.center)
    .. controls (-1.2,2.6) and (0.5,2.3) .. (Uxip.center)
    .. controls (3.2,2.6) and (4.0,2.9) .. (4.5,2.7);
\node[below=3pt, black!80, font=\small] at (4.6,2.7) {section $\sigma$};

\draw[green!50!black, thick, ->] 
    (Uxi) .. controls (-0.5,4.8) and (0.8,4.9) .. (gUxi);
\node[green!50!black, font=\normalsize, above] at (-0.2,4.7) {$g \in G$};

\draw[orange!80!black, thick, ->] (gUxi) -- (Uxip);
\node[left=3pt, orange!80!black, font=\normalsize] at ($(gUxi)!0.5!(Uxip)$) {$h^{-1}$};

\end{tikzpicture}
        }
    \caption{Schematic illustration of the coset relation in \cref{eqnSM: gU=U'h}. The total space $G$ is a principal $H$-bundle over $G/H$. The left action by $g$ maps the representative $U(\goldstone)$ to a new element $gU(\goldstone)$ in the fiber over $\goldstone'$, which differs from the chosen representative $U(\goldstone')$ by an $H$-element $h(\goldstone,g)$ along the fiber.}
    \label{figSM: coset fiber}
\end{figure}
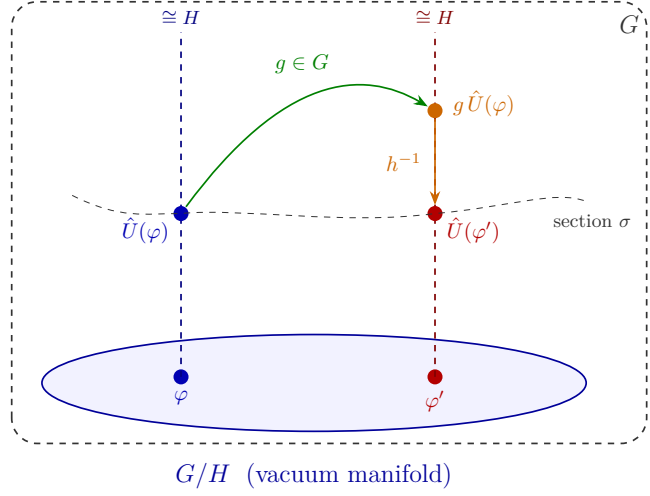

A natural transformation to consider is the left action by a group element $g \in G$ on the coset representative, mapping $\hU(\goldstone)$ to $g \hU(\goldstone)$. 
The product $g\, \hU(\goldstone)$ is again an element of $G$, and therefore belongs to some left coset in $G/H$. We may label this new coset by a new set of Goldstone coordinates $\goldstone'$.
However, $g\, \hU(\goldstone)$ will not, in general, coincide with the particular representative $\hU(\goldstone')$ chosen for that coset. Rather, the two can differ by right multiplication by an element of the unbroken subgroup $H$:
\begin{equation}
\label{eqnSM: gU=U'h}
    g \, \hU(\goldstone) = \hU(\goldstone') \, h, \quad h(\goldstone, g)\in H.
\end{equation}
The element $h(g,\goldstone)$ is the compensator: it restores the transformed group element to the chosen coset section, where the redundant $H$-coordinates have been set to zero. This relation defines the nonlinear transformation law of the Goldstone fields, $\goldstone \rightarrow \goldstone'(g,\goldstone)$, as illustrated in \cref{figSM: coset fiber}.

To construct the effective Lagrangian, we need a quantity built from the coset representative $\hU(\goldstone)$ that has a manageable transformation law under $G$. 
The basic object is the MC form
\begin{equation}
    \gamma = -i\,\hU^{-1}\diffD \hU ,
\end{equation}
where $\diffD$ acts on the Lie manifold. 
Under a $G$-transformation, the coset representative transforms as
\begin{equation}
    \hU(\goldstone') = g \, \hU(\goldstone) \, h^{-1}(g,\goldstone)
\end{equation}
Since $g$ is a constant, the MC form transforms as
\begin{equation}
\label{eqnSM: U dU transform under h}
    \gamma' = h\gamma h^{-1} - i h \, \diffD h^{-1}
\end{equation}
Since the Lie algebra admits a vector space decomposition $\mathfrak{g} = \mathfrak{h} \oplus \mathfrak{m}$, the MC form decomposes as
\begin{equation}
    \gamma 
    = -i \hU^{-1} \diffD \hU 
    = \Omega + \mathcal{A}
    = \sum_m \Omega^{m} \hbroken_m + \sum_A \mathcal{A}^{A} \hT_A
    , \quad \Omega \in \mathfrak{m}~
    , \quad \mathcal{A} \in \mathfrak{h} ~.
\end{equation}
For a reductive coset, $[\mathfrak h,\mathfrak m]\subset \mathfrak m$, the two components transform separately as
\begin{equation}
\label{eqnSM: coset Omega A transform}
    \Omega \;\mapsto\; h\, \Omega \, h^{-1} ~
    , \quad 
    \mathcal{A} \;\mapsto\; h\, \mathcal{A} \, h^{-1} -i h\, \diffD h^{-1} ~.
\end{equation}
Thus the broken component $\Omega$ transforms covariantly under the unbroken group $H$, and its components $\Omega^i$ provide covariant building blocks for the effective Lagrangian. 
By contrast, the unbroken component $\mathcal{A}$ transforms as a connection on a principal $H$-bundle. It can therefore be used to define covariant derivatives of $H$-covariant objects. For example, in Lie-algebra notation,
\begin{equation}
    \text{D} \Omega 
    = \diffD \Omega + i [ \mathcal{A}, \Omega ]_{\wedge}
    = \diffD \Omega + i ( \mathcal{A} \wedge \Omega + \Omega \wedge \mathcal{A} )
.\end{equation}
Using the transformation laws in \cref{eqnSM: coset Omega A transform}, one can show that $\text{D}\Omega \mapsto h \text{D}\Omega h^{-1}$. Thus $\text{D}\Omega$ transforms covariantly, just like $\Omega$ itself. Similarly, the field strength two-form 
\begin{equation}
    \mathcal{F} = \diffD \mathcal{A} + i \mathcal{A} \wedge \mathcal{A}
\end{equation}
also transforms covariantly as $\mathcal{F} \mapsto h \mathcal{F} h^{-1}$ under \cref{eqnSM: coset Omega A transform}. To construct an $H$-invariant Lagrangian, broken indices are contracted with $H$-invariant tensors. We denote by $\kappa_{mn}$ an $H$-invariant metric on $\mathfrak m$, satisfying
\begin{equation}
    \mathcal{R}_{m}{}^{m'}(h) \mathcal{R}_{n}{}^{n'}(h) \kappa_{m'n'} = \kappa_{mn}
    ,\quad
    h \hbroken_m h^{-1} = \mathcal{R}_{m}{}^{n}(h) \hbroken_n
.\end{equation}

We now introduce spacetime coordinates $x^\mu$, where $\mu=0, 1, \ldots, d$, and promote the Goldstone coordinates to fields: $\goldstone^m \rightarrow \goldstone^m(x)$.
The MC form is then pulled back to spacetime:
\begin{equation}
    \Omega
    =
    \Omega^m \hbroken_m
    =
    \diffD x^\mu\, \Omega_\mu^m(x)\,\hbroken_m ,
    \qquad
    \mathcal A
    =
    \mathcal A^A \hT_A
    =
    \diffD x^\mu\, \mathcal A_\mu^A(x)\,\hT_A .
\end{equation}
The effective Lagrangian is obtained by contracting the broken indices with $H$-invariant tensor $\kappa_{mn}$, and contracting the spacetime indices with tensors compatible with the spacetime symmetries $g_{\mu\nu}$.
The leading two-derivative term takes the form
\begin{equation}
    L \sim g^{\mu\nu} \kappa_{mn} \Omega_{\mu}^{m} \Omega_{\nu}^{n}
,\end{equation}
while higher-derivative operators give corrections suppressed by powers of the EFT cutoff scale.

\subsection{The Heisenberg model: spin wave}

As a simple example, consider an internal spin-rotation symmetry broken as $SO(3)\rightarrow SO(2)$. Choose the reference GS $\ket{\GS_0}$ to be invariant under the unbroken generator $\hat{S}_z$. The broken generators are then $\hat{S}_x$ and $\hat{S}_y$, so a convenient coset representative is
\begin{equation}
    \hU(\goldstone_x, \goldstone_y) 
    = \exp \left( i \goldstone_x \hat{S}_x + i \goldstone_y \hat{S}_y \right)
.\end{equation}
The MC form is
\begin{equation}
\begin{aligned}
    -i \hU^{-1} \partial_\mu \hU 
    &= 
    \sum_{a, b = x, y} \left[ \frac{\sin\rho}{\rho} \partial_\mu \goldstone_a + \left( 1 - \frac{\sin\rho}{\rho} \right) \frac{\goldstone_a \goldstone_b}{\rho^2} \partial_\mu \goldstone_b  \right] \hat{S}_a
    + \frac{1-\cos\rho}{\rho^2} \left( \goldstone_x \partial_\mu \goldstone_y - \goldstone_y \partial_\mu \goldstone_x \right) \hat{S}_z,
    \\
    &\approx
    (\partial_\mu \goldstone_x) \hat{S}_x
    + (\partial_\mu \goldstone_y) \hat{S}_y
    + \frac{1}{2} \left( \goldstone_x \partial_\mu \goldstone_y - \goldstone_y \partial_\mu \goldstone_x \right) \hat{S}_z
,\end{aligned}
\end{equation}
where $\rho \equiv \sqrt{\goldstone_x^2 + \goldstone_y^2}$. The components along $\hat{S}_x$ and $\hat{S}_y$ are the covariant derivatives of the Goldstone fields, while the component along $\hat{S}_z$ is the unbroken $SO(2)$ connection. Since the invariant metric on the broken subspace is $\kappa_{ij}=\delta_{ij}$, the leading two-derivative invariant is
\begin{equation}
    \mathcal{L}
    \sim (\partial_\mu \goldstone_x)^2 
    + (\partial_\mu \goldstone_y)^2
,\end{equation}
Equivalently, one may introduce the unit vector order parameter $\vec n$ by
\begin{equation}
    \vec{n} \cdot \vec{S} = \hU \hat{S}_z \hU^{-1}
.\end{equation}
For the above parameterization,
\begin{equation}
    \vec{n} 
    = \left( 
    -\frac{\sin\rho}{\rho}\goldstone_y, 
    \frac{\sin\rho}{\rho} \goldstone_x , \cos\rho \right) 
    \approx \left( -\goldstone_y, \goldstone_x, 1 \right)
.\end{equation}
The leading invariant can then be written as
\begin{equation}
    \mathcal{L} \sim (\partial_\mu \vec{n})^2
.\end{equation}

\subsection{With spacetime symmetry}

When spacetime symmetries are involved, or when the broken internal generators do not commute with translations, it is useful to include translation generators ($\hP_a$ along the $a$-th direction) explicitly in the coset representative, where $\hbroken_m$ are the broken non-translation generators:
\begin{equation}
    \hU(X,\tilde{\goldstone})
    =
    e^{-iX^a\hP_a} e^{i\tilde{\goldstone}^m\hbroken_m}.
\end{equation}
Here $X^a$ is a coordinate along the translation directions of the enlarged parametrization, not yet a physical spacetime coordinate. To be generic, here we assume $\hP_a$ runs over both spatial and time directions, with the time-direction translation operator being simply the Hamiltonian $H$. Since $H$ commutes with all the other symmetry generators, including $H$ as a generator or not does not affect our discussions hereafter.
Note that we intentionally place $e^{-iX^a\hP_a}$ on the left regardless whether $\hP_a$ is broken or not. This ensures that the translations $e^{i\epsilon^a\hP_a}$ act manifestly on the coset representative as changing $X^a\to X^a-\epsilon^a$. This also ensures the Goldstone action we obtain later to be translationally invariant, if the translation symmetry (or a dressed version of it) is unbroken. 
Other orderings are equivalent coordinate choices, but are less transparent when $[\hP_a,\hbroken_m]\neq0$.

If translations are unbroken for the reference state (which is the case we considered in our work), then
\begin{equation}
    e^{ia^a\hP_a}\ket{\GS_0}
    =
    \ket{\GS_0},
\end{equation}
up to the usual identification of physically equivalent states. Thus the physical GS manifold is still labeled only by the Goldstone coordinates $\tilde{\goldstone}^i$. The enlarged representative therefore contains an overcomplete parametrization $(X,\tilde{\goldstone})$, giving the equivalence
\begin{equation}
    \hU(X,\tilde{\goldstone})
    \sim
    \hU(X,\tilde{\goldstone})e^{i\lambda^a\hP_a}\ .
\end{equation}
Using the algebra of translations with the broken generators, this equivalence can be rewritten as an equivalence condition of the coordinates $(X,\tilde{\goldstone})$. In the
simple case where translations commute among themselves and act linearly on the broken generators as
\begin{equation}
    [\hP_a,\hbroken_m]
    =
    -i(A_a)_m^n\hbroken_n ,
    \quad
    \text{or, equivalently,}
    \quad
    [\hP_a, \bold{\hbroken}]
    =
    -iA_a \bold{\hbroken} ,
\end{equation}
one finds
\begin{equation}
    e^{-iX^a\hP_a}e^{i\tilde{\goldstoneSet}^{T} \bold{\hbroken}}e^{i\lambda^a\hP_a}
    =
    e^{-i(X^a-\lambda^a)\hP_a}
    \exp\left[ i \tilde{\goldstoneSet}^{T} e^{-\lambda^a A_a}\bold{\hbroken} \right].
\end{equation}
Thus, for $\ket{\GS_0}$,
\begin{equation}
\label{eqnSM: redundant X phi transformation}
    (X,\tilde{\goldstoneSet})
    \sim
    (X-\lambda,e^{-\lambda^a A_a^{T}}\tilde{\goldstoneSet})
.\end{equation}
More general algebras may produce additional BCH corrections, but the essential point is that a shift in the translation coordinate $X^a$ can be compensated by a corresponding transformation of the Goldstone coordinates.

After promoting the coordinates $(X^a, \tilde\goldstoneSet)$ to fields depending on auxiliary coordinate $\sigma^a$, not yet identified with physical spacetime coordinates, $(X^a, \tilde{\goldstoneSet}) \to (X^a(\sigma), \tilde{\goldstoneSet}(\sigma))$, one may impose the equivalence in \cref{eqnSM: redundant X phi transformation} pointwise by promoting $\lambda^a$ to an arbitrary function of $\sigma$: $\lambda^a\to \lambda^a(\sigma)$.
The corresponding local redundancy is
\begin{equation}
    X^a(\sigma)\to X^a(\sigma)-\lambda^a(\sigma),
    \qquad
    \tilde{\goldstoneSet}(\sigma)\to
    e^{-\lambda^a(\sigma)A_a^{T}}\tilde{\goldstoneSet}(\sigma) .
\end{equation}
This transformation is not an ordinary physical translation symmetry. Rather,
it is a gauge redundancy introduced by the enlarged parametrization. 
The redundancy may be fixed, at least locally, by choosing $\lambda^a(\sigma)=X^a(\sigma)-\sigma^a$, so that
\begin{equation}
    X^a(\sigma)=\sigma^a \equiv x^a 
    ,\quad 
    \goldstoneSet(\sigma) = e^{-\lambda^a(\sigma)A_a^{T}}\tilde{\goldstoneSet}(\sigma)
    .
\end{equation}
Only after this gauge choice is the auxiliary coordinate $\sigma^a$ identified
with the physical spacetime coordinate $x^a$. The Goldstone fields
$\goldstoneSet(\sigma)$ then become ordinary spacetime fields
$\goldstoneSet(x)$, and the coset representative reduces to
\begin{equation}
    \hU(x,\goldstone)
    =
    e^{-ix^a\hP_a}e^{i\goldstoneSet^{T}(x)\bold{\hbroken}}.
\end{equation}
A left action by $g=e^{i\epsilon^a\hP_a}$ gives
\begin{equation}
    g\hU(x,\goldstone(x))
    =
    \hU(x-\epsilon,\goldstone(x)).
\end{equation}
Equivalently, if the transformed field is written as a function of the translated coordinate, then $\goldstone'(x-\epsilon)=\goldstone(x)$.

The MC form then decomposes into components along translations, broken generators, and unbroken generators:
\begin{equation}
    \gamma 
    = -i\,\hU^{-1} \diffD \hU
    =
    -e^a \hP_a
    +
    \Omega^m \hbroken_m
    +
    \mathcal A^A \hT_A 
.\end{equation} 
Since the translation generators $\hP_a$ transform as local-frame vectors under the unbroken spacetime rotations, the corresponding MC component $e^a$ is a covariant one-form. It is therefore identified with the vielbein.
The vielbein must also be included in the effective action. For example, a two-derivative term takes the form
\begin{equation}
    \mathcal{L} \supset \sqrt{-\det(g)}\; \kappa_{mn}\, g^{\mu\nu}\, \Omega_\mu^m\, \Omega_\nu^n ~,
\end{equation}
where $g^{\mu\nu}$ is the inverse of the metric $g_{\mu\nu} = e_\mu^a\, e_\nu^b\, \eta_{ab}$, and $\mu,\nu$ runs over all spacetime components. When the vielbein is trivial, $e_\mu^a = \delta_\mu^a$ (which is assumed in the main text), the translation operator satisfies $\hP_\mu = e_\mu^a \hP_a = \hP_a$, and the Lagrangian reduces to $\mathcal{L} \supset \kappa_{mn}\, \eta^{\mu\nu}\, \Omega_\mu^m\, \Omega_\nu^n$.

\section{Applying the coset construction to TCMS}

In the main text, we introduced the Goldstone fields as parameters of the order parameter and constructed a generic Lagrangian for their dynamics after TCMS is broken. In this section, we rederive the low-energy effective theory using the coset construction, and show that it has the same general form.
We work in flat space, where coordinate and local-frame indices may be identified. Thus we do not distinguish $\mu,\nu$ from $a,b$ explicitly, and write $\hP_\mu=\hP_a$.
In addition, since we only consider TCMS in the spatial directions, $\mu$ runs only over the spatial coordinates, $\mu=1, \dots, d$.
The coset is parameterized by:
\begin{equation}
\label{eqn: coset parameterization}
    \hU(\boldx, \goldstoneSet) = e^{-i \hP_{\mu} x^{\mu}} e^{i \goldstoneSet^{T} \overline{\hQset}}
\end{equation}
where $\goldstoneSet(\boldx)$ denotes the collection of Goldstone fields associated with the broken MS generated by $\overline{\hQset}$.
The coefficient of $\hP_\mu$ is $1$ because we are working in flat space with a trivial vielbein.
The transformation laws of the Goldstone fields under the symmetry $g$ can be deduced from 
\begin{equation}
g \hU(\boldx, \goldstoneSet) = \hU(\boldx', \goldstoneSet') h(g,\goldstoneSet)
.\end{equation} 
Choosing $g=e^{i\boldsymbol{\xi}^{T} \overline{\hQset}}$, we have:
\begin{equation}
\label{eqn: goldstone transformation law}
    g \hU(\boldx, \goldstoneSet) = e^{-i\hP_{\mu} x^{\mu}} \exp\left( i \boldsymbol{\xi}^{T} e^{\overline{A}_{\mu}^T x^{\mu}}  \overline{\hQset} \right) e^{i \goldstoneSet^{T} \overline{\hQset}} h
    ,\qquad
    \goldstoneSet' = \goldstoneSet + e^{\overline{A}_{\mu}^T x^{\mu}} \boldsymbol{\xi},
\end{equation}
which shows the transformation law of the Goldstone fields.
Here, $h$ is an element of the unbroken subgroup generated by $\hQset'$.
To build objects that are invariant under the total $g$ transformation, we look at the MC form, which can be expanded on the basis of generators as:
\begin{equation}
\label{eqn: general eq of U dU}
    -i \hU^{-1} \partial_{\mu} \hU 
    = -\hP_{\mu} + \boldsymbol{\Omega}_{\mu}^{T} \overline{\hQset}
\end{equation}
where $\boldsymbol{\Omega}_\mu = (\partial_{\mu} - \overline{A}_{\mu}^{T}) \goldstoneSet$ is the covariant derivative of the Goldstone fields, which transform homogeneously under the unbroken group, and therefore serve as the natural building blocks for the most general invariant effective Lagrangian. The most general quadratic effective Lagrangian takes the form
\begin{equation}
    \mathcal{L} = 
    \frac{1}{2} \dot{\goldstoneSet}^{T} M \dot{\goldstoneSet} 
    - \frac{1}{2} \boldsymbol{\Omega}_{\mu}^{T} W^{\mu\nu} \boldsymbol{\Omega}_{\nu}\ ,
\end{equation}
which agrees with the main text result.

\section{The inverse Higgs constraint}

The effective Lagrangian with broken monopole and dipole symmetries is constructed from the MC components $\Omega_{\text{m}}$ and $\Omega_{\text{d}}$:
\begin{equation}
    \mathcal{L} 
    = \frac{1}{2}\dot{\goldstone}_{\text{m}}^2 + \frac{1}{2}\dot{\goldstone}_{\text{d}}^2
    - \frac{1}{2} v_{\text{m}}^2 (\underbrace{\partial_x \goldstone_{\text{m}} - \goldstone_{\text{d}}}_{\Omega_{\text{m}}})^2
    - \frac{1}{2} v_{\text{d}}^2 (\underbrace{\partial_x\goldstone_{\text{d}}}_{\Omega_{\text{d}}})^2
\end{equation}
Solving the quadratic theory gives one gapped mode and one gapless mode with quadratic dispersion, $E\sim k^2$. The same low-energy result can be obtained directly by imposing the inverse Higgs constraint (IHC).

The IHC is applicable when having redundant Goldstone fields: breaking $\hQmul{\text{m}}$ forces $\hQmul{\text{d}}$ to be broken (due to $[\hP, \hQmul{\text{d}}]=-i \hQmul{\text{m}}$), so there should be effectively one low-energy degree of freedom. One would have to impose the constraint by requiring the MC component along $\hQmul{\text{m}}$ to vanish: $\goldstone_{\text{d}} = \partial\goldstone_{\text{m}}$, and the Lagrangian becomes (keeping only the lowest orders):
\begin{equation}
    \mathcal{L} = \frac{1}{2}\dot{\goldstone}_\text{m}^2 - \frac{v_\text{d}^2}{2} (\partial_x^2 \goldstone_\text{m})^2
\end{equation}
This is consistent with the low-energy energy spectrum $E= v_\text{d} k^2$ solved from the effective Lagrangian in the main text by dropping out the gapped excitations. Therefore, imposing inverse Higgs constraint at the level of effective Lagrangian is a convenient way to directly integrate out the gapped high energy excitations without changing the low energy spectrum.
Note that although there are two broken symmetries, there is only one Goldstone mode. The mismatch of the counting is common in broken symmetries involving spacetime symmetries \cite{Watanabe_2013, Watanabe_2020}. For more on the inverse Higgs constraints in a more general setup $[\hP,\overline{\hQ}^{(m)}]\supset \overline{\hQ}^{(n)}$, we refer the reader to Ref.~\cite{Brauner_2014,Ivanov:1975zq,Nicolis:2015sra} for more details.

\section{Explicit calculation of the dispersion relation}

In this section, we explicitly expand the dispersion relations for the examples discussed in the main text. For the general model, the dispersion relation is given by
\begin{equation}
\left[-\omega^2 M + (k_\mu-i\overline{A}_\mu) W^{\mu\nu}(k_\nu +i \overline{A}_\nu^T)\right]\tilde{\goldstoneSet}_{\boldsymbol{k},\omega}=0
.\end{equation}
In the following, we focus on two cases: $A_1=\JordanReal_2(0)$, corresponding to the phase that breaks all symmetries of the dipole symmetry model, and $A_1=\JordanComplex_1(0,\beta)$, corresponding to the harmonic symmetry model.

\subsection{The dipole symmetry model}

When both the dipole and monopole symmetries are broken, and the Goldstone Lagrangian is taken, for example, to be specified by $M=\mathds{1}$, $W^{11}=\text{diag}(v_\text{d}^2, v_\text{m}^2)$, and $W^{\mu\nu}=\delta_{\mu\nu}\mathds{1}$ for all $\mu,\nu>1$, the dispersion relation is:
\begin{equation}
\begin{aligned}
    &\det
    \begin{pmatrix}
        -\omega^2 + v_\text{m}^2 + v_\text{d}^2 k_1^2 + k_\perp^2 &
        - i v_\text{m}^2 k_1 \\
        i v_\text{m}^2 k_1 &
        -\omega^2 + v_\text{m}^2 k_1^2 + k_\perp^2
    \end{pmatrix}
    =0
    \\
    \Rightarrow &\;
    \omega^4 
    - \underbrace{\left[(v_\text{d}^2 + v_\text{m}^2) k_1^2 + v_\text{m}^2 + 2 k_\perp^2 \right]}_{\mathcal{B}} \omega^2 
    + \underbrace{\left[ v_\text{d}^2 v_\text{m}^2 k_1^4 + (v_\text{d}^2 + v_\text{m}^2) k_1^2 k_\perp^2 + v_\text{m}^2 k_\perp^2 + k_\perp^4 \right]}_{\mathcal{C}}
    = 0
.\end{aligned}
\end{equation}
Solving the quadratic equation for $\omega^2$ gives
\begin{equation}
    \omega_\pm^2
    =
    \frac{1}{2}
    \left[
    \mathcal{B}
    \pm
    \sqrt{\mathcal{B}^2-4\mathcal{C}}
    \right]
    = \frac{1}{2}\left[ (v_\text{d}^2 + v_\text{m}^2) k_1^2 + v_\text{m}^2 + 2 k_\perp^2 \pm \left( v_\text{m}^2 + (v_\text{d}^2 + v_\text{m}^2)k_1^2 - 2v_\text{d}^2 k_1^4 \right) \right] + \mathcal{O}(k_1^6)
,
\end{equation}
where we expand the square root at small $k_1$.
Therefore, $\omega_{-}^{2}=v_\text{d}^2 k_1^4+k_\perp ^2$ and $\omega_{+}^{2}=(v_\text{d}^2 + v_\text{m}^2) k_1^2 + v_\text{m}^2 + k_\perp^2$.
Thus, to leading order in small momentum, the gapless mode has the anisotropic
dispersion 
\begin{equation}
    \omega_-
    =
    \sqrt{v_\text{d}^2k_1^4+k_\perp^2},
\end{equation}
while the gapped mode expands as
\begin{equation}
\omega_+=v_\text{m}+\frac{(v_\text{d}^2+v_\text{m}^2)k_1^2+k_\perp^2}{2v_\text{m}}
.\end{equation}

\subsection{The harmonic symmetry model}

When the symmetries $\hQsin$ and $\hQcos$ are broken, one may take $M=\mathds{1}$, $W^{11}=v_\text{h}^2\mathds{1}$, and $W^{\mu\nu}=\delta_{\mu\nu}\mathds{1}$ for all $\mu,\nu>1$ as a representative Goldstone Lagrangian. The dispersion relation is:
\begin{equation}
\begin{aligned}
    &\det
    \begin{pmatrix}
        -\omega^2 + v_\text{h}^2(k_1^2 + \beta^2) + k_\perp^2 &
        -2i v_\text{h}^2 \beta k_1 \\
        2i v_\text{h}^2 \beta k_1 &
        -\omega^2 + v_\text{h}^2(k_1^2 + \beta^2) + k_\perp^2
    \end{pmatrix}
    =0
    \\
    \Rightarrow &\;
    \left[-\omega^2 + v_\text{h}^2 (k_1^2 + \beta^2) + k_\perp^2 \right]^2 
    = 4 v_\text{h}^4 \beta^2 k_1^2
    \\
    \Rightarrow &\;
    \omega^2 = v_\text{h}^2 (k_1 \pm \beta)^2 + k_\perp^2
\end{aligned}
\end{equation}

\section{Exponential symmetry on a lattice with periodic boundary condition}

In this section, we discuss the exponential symmetry on a periodic lattice of length $L$. We label the sites by $j \in [1,L]$ with $\hb_j \equiv \hb_{j+L}$. The local interaction is $\hat{K}_j = \hb_j (\hb_{j+1}^\dagger)^{r}$, with $r = e^{-\alpha}$.
Assuming $r \in \mathbb{Z}_{>1}$, it is convenient to define the integer-normalized charge $\tilde{Q}_{\text{e}}=\sum_{j=1}^L r^{L - j} \hn_j$. This is different from $\hQ_{\text{e}}=\sum_{j=1}^L e^{\alpha j} \hn_j$ only by an overall normalization.

Under the transformation $e^{i \xi \tilde{Q}_{\text{e}}}$, the bulk terms $\hat K_j$ ($j=1,\ldots,L-1$) are invariant. The only nontrivial constraint comes from the boundary term $\hat K_L = \hb_L (\hb_{1}^\dagger)^{r}$, which acquires the phase $\exp[i\xi(r^L-1)]$. Therefore, the boundary term remains invariant only if 
\begin{equation}
\label{eqnSM: xi in terms of integer}
    \xi = \frac{2\pi n}{r^L-1} ,
    \qquad n\in\mathbb Z .
\end{equation}
Thus, on a periodic chain, the exponential symmetry is reduced to the finite cyclic group $\mathbb{Z}_{r^L-1}$.
For the deformed Bose-Hubbard model we considered in this paper, $r=2$, so the finite symmetry group is $\mathbb{Z}_{2^L-1}$.

We now show that lattice translation acts as an automorphism of this finite symmetry group. Let $\hT \hb_j \hT^{-1} = \hb_{j+1}$, then
\begin{equation}
\hT \tilde Q_{\text{e}}\hT^{-1}
= r\tilde Q_{\text{e}} - (r^L-1)\hn_1 .
\end{equation}
Using \cref{eqnSM: xi in terms of integer} and the fact that $\hn_1$ has integer eigenvalues, we obtain
\begin{equation}
    \hT e^{i \xi \tilde{Q}_{\text{e}}} \hT^{-1} = e^{i \xi \hT \tilde{Q}_{\text{e}} \hT^{-1}} = e^{i\xi r \tilde{Q}_{\text{e}}} ~.
\end{equation}
Equivalently, translation induces the map
\begin{equation}
\label{eq:auto_map}
    \xi = \frac{2\pi n}{r^L -1} \mapsto \frac{2\pi n r}{r^L-1} ~,
\end{equation}
or $n \mapsto rn \text{ mod } (r^L-1)$.
Since ${\rm gcd}(r,r^L-1)=1$, the map is an automorphism of $\mathbb{Z}_{r^L-1}$. 

Finally, there is no monodromy after translating around the full ring. Indeed,
\begin{equation}
    \hT^L e^{i \xi \tilde{Q}_{\text{e}}} \hT^{-L} 
    = e^{i\xi r^L \tilde{Q}_{\text{e}}} 
    = e^{i\xi \tilde{Q}_{\text{e}}},
\end{equation}
because $r^L \equiv 1 \text{ mod }(r^L-1)$. Therefore the finite exponential symmetry is fully compatible with periodic boundary conditions and lattice translation.

\end{document}